\documentclass[12pt]{article}
\usepackage{amssymb}
\usepackage{bm}
\usepackage{a4}
\usepackage{amsthm}
\usepackage{amsmath}
\usepackage{graphicx}
\evensidemargin \oddsidemargin
\newcommand{\CC}{\mathbb{C}}
\newcommand{\RR}{\mathbb{R}}
\newcommand{\ZZ}{\mathbb{Z}}
\newcommand{\NN}{\mathbb{N}}

\newcommand{\Hi}{{\cal H}}
\newcommand{\Ob}{{\cal O}}

\newcommand{\A}{{\cal A}}
\newcommand{\C}{{\cal C}}
\newcommand{\R}{{\cal R}}
\newcommand{\bv}{{\bf v}}
\newcommand{\bx}{{\bf x}}

\newcommand{\bpi}{{\bm\pi}}

\newcommand{\be}{\begin{equation}}
\newcommand{\ee}{\end{equation}}
\newcommand{\bea}{\begin{eqnarray}}
\newcommand{\eea}{\end{eqnarray}}
\newcommand{\ba}{\begin{array}}
\newcommand{\ea}{\end{array}}
\def\nn{\nonumber \\}

\newtheorem{propo}{Proposition}[section]

\begin{document}
\title{Fuzzy circle and new fuzzy sphere through confining potentials and energy cutoffs}
\date{}

\author{  Gaetano Fiore, Francesco Pisacane
   \\   \\  
Dip. di Matematica e applicazioni, Universit\`a di Napoli ``Federico II'',\\
\& INFN, Sezione di Napoli, \\
Complesso Universitario  M. S. Angelo, Via Cintia, 80126 Napoli, Italy}

\maketitle

\begin{abstract}
\noindent

Guided by ordinary quantum mechanics we introduce new fuzzy spheres $S^d_{\Lambda}$
of dimensions $d=1,2$: we consider an ordinary quantum particle in $D=d+1$ dimensions subject to   
a rotation invariant potential well $V(r)$  with a very sharp minimum on a sphere of unit radius. 
Imposing a sufficiently low energy cutoff to `freeze' the radial excitations makes only a finite-dimensional Hilbert subspace 
accessible and  on it the coordinates noncommutative {\it \`a la Snyder}; in fact, on it they
generate the whole algebra of observables. The construction is equivariant not
only under rotations - as  Madore's fuzzy sphere -, but under the full orthogonal group $O(D)$. 
Making the cutoff and the depth of the well dependent 
on (and diverging with) a natural number $\Lambda$, and keeping the leading terms in $1/\Lambda$,
we obtain a sequence $S^d_{\Lambda}$  of fuzzy spheres converging  
to the sphere $S^d$ 
 in the limit $\Lambda\to\infty$ (whereby we recover ordinary quantum mechanics on $S^d$).  
These models may be useful
in condensed matter problems where particles are confined on a sphere by an  
(at least approximately) rotation-invariant  potential,
beside being suggestive of analogous mechanisms 
in quantum field theory or  quantum geometry.
\end{abstract}

\section{Introduction}

 In 1947  Snyder proposed \cite{Snyder} the first example of noncommutative spacetime 
 hoping that  nontrivial (but Poincar\'e equivariant)  commutation relations
among the coordinates,   acting as a
fundamental regularization procedure,  could 
cure  ultraviolet (UV) divergencies in quantum field theory (QFT)\footnote{The idea had originated 
 in the '30s from Heisenberg, who proposed it in a letter to Peierls
\cite{Heisenberg30}; the idea propagated via Pauli to
Oppenheimer, who asked  his student  Snyder to develop it.}. He dubbed
as {\it distasteful arbitrary} the UV regularization based on
momentum (or equivalently energy) cutoff, which had just been proposed in the literature at the time,
presumably as it broke manifest Lorentz equivariance and looked {\it ad hoc}. 
Ironically, shortly afterwards this and other
more sophisticated regularization procedures found widespread application  within 
the renormalization method;  as known, the latter has proved to be extremely successful in extracting physically
correct predictions from quantum electrodynamics, chromodynamics, and more generally
the Standard Model of elementary particle physics. 
The proposal of Snyder was thus
almost forgotten for decades (exceptions are e.g. \cite{Kad62,Mir67}).  
On the other hand, it is believed that any consistent quantum theory of gravitation will set
fundamental bounds  of the order of Planck length $l_p=\sqrt{\hbar  G/c^3}\sim 10^{-33}$cm
 on the accuracy $\Delta x$ of localization measurements.  The arguments, which in 
qualitative form go  back at least to 
\cite{Wee57,Des57,Mea64}, 
are based on a cutoff on the concentration of energy\footnote{In fact, by Heisenberg uncertainty relations
to reduce the uncertainty $\Delta x$ of the coordinate $x$ of an event one must  increase
the uncertainty  $\Delta p_x$ of the conjugated momentum component by use of high energy probes;
but by general relativity the associated concentration of energy in a small region  would produce 
a trapping surface (event horizon of a black hole) if it were too large; hence the size of this region,
and $\Delta x$ itself, cannot be lower than the associated Schwarzschild radius, $l_p$.}; they
were made more precise and quantitative by 
 Doplicher, Fredenhagen, Roberts \cite{DopFreRob95}, who also  proposed that such a bound 
could follow from appropriate noncommuting coordinates (for a  review of more recent developments
see e.g. \cite{BahDopMorPia15}).
More generally, Connes' Noncommutative Geometry framework \cite{Connes} 
allows not only to  replace the commutative algebra $\A$ of functions on a manifold $M$ by a noncommutative one, 
but also to develop on it the whole machinery of differential geometry \cite{Connes,Madore99}. 
Often one deals with a family of noncommutative deformations $\A_\lambda$ of $\A$
that become commutative in some limit of the family's ruling parameter(s) $\lambda$, exactly
as the algebra of observables in ordinary quantum mechanics becomes 
the algebra of functions on phase space  as $\hbar\to 0$. 

Fuzzy spaces are particular examples parametrized by a positive integer $n$, so that the 
algebra ${\cal A}_n$ is a  finite-dimensional matrix algebra with dimension which increases and diverges with $n$ while $\A_n\to \A$ (in a suitable sense). Since their introduction they have raised big interest in the high energy physics community as a non-perturbative technique in  QFT (or string theory) based on a finite-discretization 
of space(time)  alternative to the lattice one: the advantage is that the algebras 
$\A_n$ can carry representations of Lie groups (not only of discrete ones).
The first and seminal fuzzy spaces are the Fuzzy Sphere (FS) of Madore and Hoppe \cite{Mad92,HopdeWNic}
 and the noncommutative torus  \cite{Rie85,ConRie85}
parametrized by a root of unity (this is often called fuzzy torus by theoretical physicists, see e.g. \cite{Kim01})\footnote{In \cite{HopdeWNic} the FS was used in connection with the quantization of the membrane; its noncommutative differential geometry
was first constructed by Madore in \cite{Mad92}. The algebra of the noncommutative torus (NT) is generated by unitary elements $U,V$ fulfilling $UV=VUq$, with $q$ on the unit circle; when $q$ is a root of unity the irreducible representations are finite-dimensional, and the NT can be dubbed as {\it fuzzy}.}; 
the first applications to QFT models of the FS are in \cite{GroMad92,GroKliPre96'}. 
The FS is a sequence of $SO(3)$-equivariant, finite noncommutative $*$-algebras ${\cal A}_n$ isomorphic to $ M_n$ (the algebra of $n\times n$ matrices);
each matrix represents  the expansion  in spherical harmonics of an element of $C(S^2)$
truncated at level $n$. 
$\A_n$ is generated by hermitean noncommutative coordinates $x^i$ fulfilling
\be
[x^i,x^j]=\frac {2i}{\sqrt{n^2-1}}\varepsilon^{ijk}x^k, \qquad
r^2:=x^ix^i=1,\qquad n\in\NN\setminus \{1\}                        \label{FS}
\ee
(here and below sum over repeated indices is understood). 
The Hilbert space is chosen as $\Hi\simeq \CC^n$ so that it carry an irreducible representation 
of \ $U\! so(3)$, and the square distance $r^2$ from the origin - which is central - be  identically equal to $1$. We note that however (\ref{FS})
are equivariant only under $SO(3)$, not $O(3)$; in particular not under parity $x^i\mapsto -x^i$. 
Fuzzy spaces can be used also in  extra dimensions to account
for  internal (e.g. gauge) degrees of freedom, see e.g. \cite{AscMadManSteZou}.

As the  arguments leading to $\Delta x\ge l_p$ suggest, imposing an energy cutoff 
$\overline{E}$ on an existing theory can be physically justified
by two reasons, at least. It may be a necessity when we believe
that $\overline{E}$  represents the threshold for the onset of new physics  not
accountable by that theory. Or it may serve to yield an effective description of the system when we, as well as  the interactions with the environment, 
are not able to bring its state to higher energies;  this leads also to 
a lower bound in the accuracy with which our apparatus
can measure some observables (position, momentum, ...) of the system, which corresponds to
the maximum energy transferable to the system by the apparatus in the measurement process, or by the envorinment during the interaction time.  (Of course, the two reasons may co-exist.)
Mathematically, the cutoff is imposed by a projection on the Hilbert subspace 
characterized by energies $E\le \overline{E}$. 
If the Hamiltonian is invariant under some symmetry group, then the projection 
is invariant as well, and the projected theory will inherit that symmetry.

That imposing such a cutoff can  modify a quantum mechanical model by converting
its commuting coordinates into non-commuting ones is simply illustrated by the
well-known  Landau model  (see e.g.  
\cite{Jackiw,Magro,KarNaiRan05,KarNai06}), 
which describes a charged quantum particle in 2D interacting only with an uniform magnetic field (in the orthogonal direction) $B$. The energy levels are $E_n=\frac{\hbar|eB|}{mc}n$, if we fix the additive constant so that the lowest level is $E_0=0$; choosing $\overline{E}\le \frac{\hbar|eB|}{mc}$ (this may be physically justified e.g. by a very strong $B$), then the Hilbert space of states  is projected to the subspace $\Hi_0$ of zero energy, and $\frac{e}{c}Bx,y$ become canonically conjugates, i.e. have a non-zero (but constant) commutator. The dimension of $\Hi_0$ is approximately proportional to the area of the surface, hence is finite (resp. infinite) if the area is.

Inspired by the projection mechanism in the Landau model, here we consider a quantum particle in 
dimension $D=2$ or $D=3$ with a Hamiltonian consisting of the standard kinetic term and a rotation invariant potential energy $V(r)$  with a very deep minimum (a well) respectively on a circle or on a sphere of unit radius;  $k\equiv V''(1)/4>0$ plays the role
of confining parameter. Imposing an energy cutoff $\overline{E}$ makes only a finite-dimensional Hilbert subspace $\Hi_{\overline{E}}$
accessible and the projected coordinates noncommutative on $\Hi_{\overline{E}}$. We choose
${\overline{E}}< 2\sqrt{2k}$ so that $\Hi_{\overline{E}}$ does not contain excited radial modes,
and on it the Hamiltonian reduces to the square angular momentum; this can be considered as a quantum version of the constraint $r=1$.
It turns out that the coordinates  generate the whole algebra $\A_{\overline{E}}:= End\big(\Hi_{\overline{E}}\big)$ of observables on $\Hi_{\overline{E}}$. Their commutators are of Snyder type, i.e. proportional to the angular momentum components $L_{ij}$ (apart from a small correction - depending only on the square angular momentum - on the highest energy states), rather than some function of the coordinates. Moreover,  $(\A_{\overline{E}},\Hi_{\overline{E}})$ is equivariant under the {\it full} group $O(D)$ of orthogonal transformations, because both the starting  quantum mechanical model on ${\cal L}^2(\RR^D)$ and the cut-off procedure are. Actually, we prove the realization $\A_{\overline{E}}=\pi_{\overline{E}}\big[U\!so(D\!+\!1)\big]$, with $\pi_{\overline{E}}$
a suitable irreducible unitary representation 
of $U\!so(D\!+\!1)$ on $\Hi_{\overline{E}}$; as a consequence,
$\Hi_{\overline{E}}$ carries a {\it reducible} representation of the subalgebra 
\ $U\!so(D)$ \ generated by the angular momentum components
$L_{ij}$, more precisely the direct sum
of all irreducible representations fulfilling $E\le\overline{E}$; in the $\overline{E}\to\infty$  limit
this becomes the decomposition of  the Hilbert space ${\cal L}^2(S^d)$, $d\!=\!D\!-\!1$. 
This welcomed property is not shared by the FS \cite{Mad92,HopdeWNic}.
Similarly,  the decomposition of the subspace $\C_{\overline{E}}\subset\A_{\overline{E}}$ 
of completely symmetrized polynomials in the noncommutative coordinates  into irreducible $U\!so(D)$-components  becomes the  decomposition of  the commutative algebra $C(S^d)$ [which acts on ${\cal L}^2(S^d)$ and has the same decomposition\footnote{In fact, spherical harmonics make up a basis for both ${\cal L}^2(S^d)$ and $C(S^d)$.}].
On  $\Hi_{\overline{E}}$ the square distance $\R^2$ from the origin is not identically 1, but
a function of the square angular momentum such that 
nevertheless its spectrum is very close to 1 and collapses to 1 in the $k\to\infty$  limit
of an infinitely narrow and deep well;  
the latter limit is automatic as we have to set $k\sim \overline{E}^2$ for consistency.
Thus the confining parameter $k$, or equivalently the energy cutoff $\overline{E}$, or a suitable natural number $\Lambda$ which we shall adopt to discretize both,
will also parametrize the noncommutativity of the coordinates.
Finally, there are natural embeddings \ $\Hi_\Lambda\hookrightarrow {\cal L}^2(S^d)$, 
$\C_\Lambda\hookrightarrow B(S^d),C(S^d)$, and
in a suitable sense \ $\Hi_\Lambda\to {\cal L}^2(S^d)$, 
$\C_\Lambda\to C(S^d)$ and
$\A_\Lambda$ goes to the whole algebra of observables on ${\cal L}^2(S^d)$,
 as $\Lambda\to\infty$.

We think that our models are interesting not only as 
new toy-models of fuzzy geometries in QFT and quantum geometry, but also
in view of potential applications to quantum models 
in condensed matter physics with an effective
 one- or two-dimensional configuration space in the form of a circle, a cylinder
 or a sphere\footnote{For instance, circular quantum waveguides, graphene nanotubes and fullerene. These are very thin wires or layers of matter where electrons are confined by potential energies with very deep minima 
there and steep gradients in the normal direction(s). The Hamiltonian on a cylinder can be written as the sum of the
one on the transverse section circle and of the kinetic term $-\hbar^2(\partial/\partial z)^2/2m$ in the direction
of the axis.},
because  they respect parity, and  the restriction to the circle, cylinder or sphere is an effective one obtained ``a posteriori'' from the exact dynamics in higher dimension. Moreover, our procedure can  be generalized in a straightforward manner
to $D>3$, as well as to other confining potentials; 
the dimension of the accessible Hilbert space $\Hi_{\overline{E}}$
 will be approximately ${\cal B}/h^D$, 
where  $h,{\cal B}$ are the Planck constant and the  volume of the 
classically allowed region in phase space
(i.e. the one characterized by 
$E\le \overline{E}$). All features of these new
fuzzy geometries deserve investigations, in particular their metric aspects, as done e.g. in \cite{DanLizMar14} for the FS.

Fuzzy spheres  based on some Snyder-type commutation relations have already
been proposed for $d=4$  in  \cite{GroKliPre96} (see e.g. also \cite{Ste16,Ste17}) and for all $d\ge 3$ in  \cite{Ramgoolam,Dolan:2003th}. 
In section \ref{Conclu} we sketch how we expect 
the results based on our approach would be related to  the latter.

The plan of the paper is as follows. In section \ref{genset} we introduce
the framework valid for any $D$. In sections \ref{D=2}, \ref{D=3} we treat 
the cases $D=2,3$ leading to $S^1_\Lambda,S^2_\Lambda$ respectively. 
Section \ref{Conclu} contains final remarks, outlook and
conclusions. In the Appendix (section \ref{Appendix}) we have concentrated lengthy computations and proofs.


\section{General setting}
\label{genset}

As said, we consider a quantum particle in $\mathbb{R}^D$ configuration space with Hamiltonian
\bea
H=-\frac 12\Delta + V(r),                                                 \label{Ham}
\eea
where $r^2:= x^ix^i$, $\Delta:= \partial_i\partial_i$ (sum over repeated indices understood), 
$\partial_i:= \partial/\partial x^i$, $i=1,...,D$; the cartesian coordinates $x^i$, the momentum components $-i\partial_i$, $H$ itself are  normalized
 so as to be dimensionless. $x^i,-i\partial_i$ generate the Heisenberg algebra $\Ob$ of observables.
The  canonical commutation relations
\bea
[x^i,x^j]=0,\qquad [-i\partial_i,-i\partial_j]=0,\qquad [x^i,-i\partial_j]= i\delta^i_j    \label{ccr}
\eea
as well as the Hamiltonian are invariant under all orthogonal transformations 
\bea
x^i\mapsto x'{}^i=Q^i_j x^j, \qquad Q^{-1}=Q^T          \label{ortho}
\eea
(including parity $Q=-I$). This implies $[H,L_{ij}]=0$, where
 $L_{ij}:=ix^j\partial_i-ix^i\partial_j$ are the angular momentum components.
We shall assume that 
$V(r)$   has a very sharp minimum at \ $r=1$ with very large $k= V''(1)/4>0$, \ and fix 
$V_0:= V(1)$ so that the ground state has zero energy, i.e. $E_0=0$. We  choose an energy cutoff $\overline{E}$
fulfilling first of all the condition
\be
V(r)\simeq V_0+2k (r-1)^2\qquad \mbox{if $r$ fulfills}\quad V(r)\le  \overline{E}
\label{cond1}
\ee
so that we can neglect terms of order higher than two in the Taylor expansion of $V(r)$ around $1$
and approximate the potential as a harmonic one  in the
classical region $v_{\overline{E}}$ compatible with the energy cutoff \ $V(r)\le \overline{E}$. \ 
We are interested in finding
the eigenfunctions of $H$ 
\be
H\psi=E\psi, \qquad\psi\in {\cal L}^2\left(\mathbb{R}^D\right)                       \label{Heigen}
\ee
with eigenvalues $E\le \overline{E}$ and restricting  quantum mechanics
to the finite-dimensional Hilbert subspace $\Hi_{\overline{E}}$ spanned by them. This means that 
we shall replace every observable $A$ by $\overline A:= P_{\overline{E}}AP_{\overline{E}}$,  where 
$P_{\overline{E}}$ is the projection on $\Hi_{\overline{E}}$, and give to $\overline A$ the same physical interpretation. 
In particular: $\overline{x}^i$ will be intepreted as the observable associated to the measurement of the $i$-th
coordinate of the particle; $\overline{H}=H$ will still appear as the Hamiltonian in the original Schr\"odinger equation.
We shall also replace 
any Schr\"odinger equation $i\frac{\partial}{\partial t}\psi=H_e\psi$, with
some extended Hamiltonian $H_e=H+H'$ (containing a ``small''  extra term $H'$ representing some additional
interaction), with the finite-dimensional one $i\frac{\partial}{\partial t}\psi=\overline{H_e}\psi$
within $\Hi_{\overline{E}}$. 

By  (\ref{cond1}),  $v_{\overline{E}}\subset \RR^D$ is approximately the shell
$|r\!-\!1|\le \sqrt{\frac{\overline{E}\!-\!V_0}{2k}}$; in the limit in which both $k,\overline{E}$ diverge, but the right-hand side 
goes to zero,  $v_{\overline{E}}$ reduces to the unit sphere $S^{D-1}$. 
We expect that
in this limit the dimension of  $\Hi_{\overline{E}}$ diverges, and we recover standard 
quantum mechanics on the sphere $S^{D-1}$. As we shall see, this is the case.

$P_{\overline{E}}$ commutes not only with $H$, but also
with the  $L_{ij}:=ix^j\partial_i-ix^i\partial_j$, which are vector fields tangent
to every sphere $r\!= $const.
The $D$ derivatives $\partial_i$ make up a globally defined basis for the linear space of smooth vector fields. The set
$B=\{\partial_r,L_{ij}\: |\: i<j\}$ ($\partial_r:=\partial/\partial r$) is an alternative complete set that is singular  for $r=0$, but globally defined 
elsewhere; for $D=2$ it is a basis, for $D>2$ it is redundant, because of the relations
\be
\varepsilon^{i_1i_2i_3....i_D}x^{i_1}L^{i_2i_3}=0.           \label{Lijrel}
\ee
This redundancy is unavoidable. In $D\!=\!3$ there are no two independent globally 
defined vector fields that are tangent to the sphere ($S^2$ is not parallelizable); for instance, the derivatives $\partial_\varphi,\partial_\theta$
with respect to the polar angles are singular at the north and south poles. One
needs all three angular momentum components, which however are constrained by
$\varepsilon^{ijk}x^{i}L^{jk}=0$.

Of course, the eigenfunctions of $H$ can be more easily determined in terms of polar coordinates $r,\varphi,...$,
recalling that the Laplacian in $D$ dimensions decomposes as follows
\be
\Delta=\partial_r^2+(D-1)\frac 1r\partial_r-\frac 1{r^2}L^2,  \label{LaplacianD}
\ee
where $\partial_r:= \partial/\partial r$ and 
$L^2=L_{ij}L_{ij}/2$ 
is the square angular momentum 
(in normalized units), i.e. the Laplacian on the sphere $S^{D-1}$. We know from the $D$-dimensional theory of angular momentum that the eigenvalues of $L^2$ are $j\left(j+D-2\right)$; then replacing the Ansatz $\psi=\tilde f(r)Y(\varphi,...)$  \ ($Y$ are eigenfunctions of $L^2$ and of the elements of a Cartan subalgebra of $so(D)$; $r,\varphi,...$ are polar coordinates) transforms the PDE \ $H\psi=E\psi$ \ into this auxiliary ODE in the unknown $\tilde f(r)$
\begin{equation}
\left[-\partial_r^2-(D-1)\frac 1r\partial_r+\frac 1{r^2}j\left(j+D-2\right)+V(r)\right]\tilde f(r)=E\tilde f(r).\label{eqpolar}
\end{equation}
If $V(r)$ keeps bounded or grows at most as $\beta/r^2$ (with some $\beta\ge 0$) as $r\to 0$, 
then in the same limit $\tilde f(r)$ vanishes as $\tilde f(r)=O\left(r^{\alpha}\right)$, with $\alpha=\sqrt{\beta+j(j+D-2)}$.
In fact, by Fuchs theorem every solution of (\ref{eqpolar}) is a combination of the two independent ones
with $r\to 0$ asymptotic behaviour $r^\alpha,r^{-\alpha}$; but the coefficient of the second must vanish in order
that $\psi\in {\cal L}^2(\RR^D)$. Hence $f(0)=0$. On the other hand,
$\psi\in {\cal L}^2(\RR^D)$ implies also   $\tilde f(r)\overset{r\rightarrow+\infty}\longrightarrow0$.
Actually, condition (\ref{cond1}) implies that
$\tilde f,\psi$ become negligibly small outside the thin shell region $V(r)\le  \overline{E}$ (around $r=1$),  and that at leading order the lowest eigenvalues $E$ are those of the $1-$dimensional harmonic oscillator approximation of (\ref{eqpolar}).

\section{D=2: $O(2)$-equivariant fuzzy circle}
\label{D=2}

We fix the notation as follows: $x\equiv x^1=r\cos{\varphi}$, $y\equiv x^2=r\sin{\varphi}$, 
$x^\pm:=(x\pm i y)/\sqrt{2}=re^{\pm i\varphi}/\sqrt{2}$; we abbreviate $u:=e^{i\varphi}$
(whence $u^\dagger=e^{-i\varphi}$),
$\partial^\mp\equiv\partial_\pm\equiv\partial/\partial x^\pm$, 
$\partial_\varphi\equiv \partial/\partial \varphi$; the angular momentum $L\equiv L_{12}$ can be expressed in the form
$L=-i\partial_\varphi=x^+\partial_+ - x^-\partial_-$, and
$$
[L,x^\pm]=\pm x^\pm, \qquad\qquad [L,\partial_\pm]=\mp \partial_\pm,
$$
i.e. the generators $x^\pm,\partial_\pm$ of the Heisenberg algebra $\Ob$ (of observables) are eigenvectors under the adjoint action
of $L$ with eigenvalues $\pm 1$. We look for $\psi$ of the form $\widetilde{\psi}_m(r,\varphi)=\widetilde{f}(r)e^{im\varphi}$,
with eigenvalues $m\in\mathbb{Z}$ and we can use the $L$-eigenvalue as a $\mathbb{Z}$-grading for both $\Hi$ and $\Ob$,
 in a compatible way:
$$
\Hi=\bigoplus_{m\in\mathbb{Z}}\Hi^m,\qquad \Ob=\bigoplus_{m\in\mathbb{Z}}\Ob^m,\qquad \Ob^m\Ob^{m'}=\Ob^{m+m'}, 
\qquad \Ob^m\Hi^{m'}=\Hi^{m+m'}
$$
(the last relation must be understood modulo domain restrictions); clearly $L,r,h(r),\partial_r,\Delta\in \Ob^0$
[for any funtion $h(r)$].
Equation (\ref{eqpolar}) becomes
$$
\widetilde{f}''(r)+\frac 1r \widetilde f'(r)+\left[E\!-\!V(r)\!-\!\frac{m^2}{r^2}\right]\widetilde{f}(r)=0.
$$
We change the radial variable
$r\mapsto \rho:=\ln r$ and set $f(\rho):=\widetilde{f}(r)=\widetilde{f}(e^\rho)$, whereby 
the  previous equation is transformed into
 $f''(\rho)\!+\!\left\{ e^{2\rho}\left[E\!-\!V(e^\rho)\right]\!-\!m^2\right\}\!f(\rho) =0$.
By condition (\ref{cond1}),  in the region $|r\!-\!1|\le \sqrt{\frac{\overline{E}\!-\!V_0}{2k}}$
we can neglect the terms of order higher than two
in the Taylor expansions \   $e^{2\rho}= 1+2\rho+2\rho^2+...$, $V(e^{\rho})= V_0+2k\rho^2+...$ \
and thus approximate  the above equation by
\bea
\ba{c}
\check H f(\rho)=e_m f(\rho), \qquad\quad
\check H:=-\partial_\rho^2+k_m\left(\rho-\widetilde{\rho}_m\right)^2,  \\[10pt]
\displaystyle\mbox{ where } \qquad k_m=2(k\!-\!E'),\quad E'=E\!-\!V_0,\quad \widetilde{\rho}_m=\frac{E'}{k_m}, \quad 
e_m=\frac{E'{}^2}{k_m}\!+\!E'\!-\!m^2,
\ea   \label{defs}
\eea
which has the form of the eigenvalue equation for a harmonic oscillator in 1 dimension with equilibrium position at
$\rho=\widetilde{\rho}_m$. In order that $\psi$ be square-integrable it must be  \ $f(\rho)\stackrel{\rho\to\infty}{\longrightarrow}0$, $f(\rho)\stackrel{\rho\to-\infty}{\longrightarrow}0$, \ 
what selects as  eigenfunctions of the auxiliary operator $\check H$ 
$$
f_{n,m}(\rho)=\exp\left[\!-\frac{\left(\rho\!-\!\widetilde{\rho}_m\right)^2\sqrt{k_m}}{2}\right] H_n\!\left[\left(\rho\!-\!\widetilde{\rho}_m\right)\sqrt[4]{k_m}\right],
$$
(here $H_n$ is the Hermite polynomial of order $n$), and as corresponding ``eigenvalues'' $e_{m,n}=(2n+1)\sqrt{k_m}$.
By (\ref{defs}) this implies that $E'$ must fulfill the equation
\bea
\frac{E'{}^2}{2(k\!-\!E')}\!+\!E'\!-\!m^2=(2n\!+\!1)\sqrt{2(k\!-\!E')}
  \label{Eneq}
\eea
Squaring both sides of (\ref{Eneq}) and multiplying them by $k_m^2$
one obtains a fourth degree equation which determines $E'$, and therefore $E$, in terms of $V_0,m,n$.
%
%
As said, we fix $V_0$ requiring that the lowest energy level, 
which corresponds to $n=m=0$, be $E_0=0$.
This implies that $V_0$ must fulfill the equation
$$
\frac{V_0^2}{2(k+V_0)}-V_0
=\sqrt{2(k+V_0)}
\qquad\Leftrightarrow\qquad
-\sqrt{\frac{1}{ 2k}}V_0-\left(\! \sqrt{\frac{1}{ 2k}}\right)^3V_0^2=\left(\!1+\frac{V_0}k\!\right)^{\frac 32}.
$$
Looking for the solution in the form $V_0=\sum_{n=-1}^\infty v_n\left(\sqrt{\frac{1}{2k}}\right)^n$ we can determine
the coefficients $v_n$, and therefore $V_0$,  solving the latter equation 
order by order in $\sqrt{\frac{1}{2k}}$. The solution is $V_0=-\sqrt{2k}+2-\frac{7}{2}\frac{1}{\sqrt{2k}}+O(1/k)$ and Figure $1$ shows the appearance of the resulting potential. Replacing this result in (\ref{Eneq}) one finds that at leading order $E$ is given by $E_{n,m}=m^2+2n\sqrt{2k}-2n+O\left(\frac{1}{\sqrt{k}}\right)$.
The term $m^2$ gives exactly what we wish, (part of) the spectrum of $L^2$ (the Laplacian on the circle).
To eliminate the subsequent, undesidered terms  we fix the energy cutoff \ 
$\overline E<2\sqrt{2k}\!-\!2$, \ so as to exclude all the states with $n>0$. 
Physically, this means that radial oscillations are ``frozen'', $n=0$,
so that all corresponding classical trajectories are circles.
The energies $E$ below $\overline E$ will therefore depend only on $m$, and will be denoted as $E_m$. Consequently, 
$k_m,\widetilde{\rho}_m$ will be determined by (\ref{defs}).
Then at leading orders in $1/\sqrt{k}$   (\ref{Eneq}) yields as eigenvalues of $L,H$ and corresponding eigenfunctions
\bea
&& L=m,\qquad H=E_m=m^2+O\left(\frac 1{\sqrt{k}}\right)
\label{Em}\\
&& \psi_m(\rho,\varphi)=N_m \: e^{im\varphi}\:
\exp\left[-\frac{\left(\rho-\widetilde{\rho}_m\right)^2\sqrt{k_m}}{2}\right],         \label{psi_m}\\[10pt]
&& \frac{k_m}{2k}=1-\frac{2}{\sqrt{2k}}+\frac{2-m^2}{k}+O\left(\frac{1}{k^{\frac{3}{2}}}\right), \qquad
\widetilde{\rho}_m=\frac{1}{\sqrt{2k}}+\frac{m^2}{2k}+O\left(\frac{1}{k^{\frac{3}{2}}}\right)  \label{krho_m}
\eea
we fix 
the  normalization factor $N_m$ so that $N_m>0$ and all  $\psi_m$ have unit norm.
The condition $E\le\overline E$ is fulfilled if we
project the theory onto the Hilbert subspace $\mathcal{H}_{{{\Lambda}}}\equiv \mathcal{H}_{\overline E}$ spanned by the $\psi_m$ with  
$|m|\leq{{\Lambda}}:=\left[\sqrt{\overline{E}}\right]$ (here $[a]$ stands for the integer part of $a>0$). 
For consistency we must choose 
\be
{{\Lambda}}^2<2\sqrt{2k} -2          \label{consistency}
\ee
so that all $E_m$ are smaller than the energy levels corresponding to $n>0$, as we can see from Figure $2$;
this is also sufficient to guarantee that $k_m
\gg 1$ for all $|m|\leq {{\Lambda}}$ \ (by the way,
$k_m>0$ is a  necessary condition for $\check H$ to be the Hamiltonian of a harmonic oscillator). 
The spectrum of $\overline{H}$ becomes the whole spectrum $\{m^2\}_{m\in \NN_0}$ of   $L^2$ 
in the limit $\Lambda,k \to\infty$ respecting (\ref{consistency}).
   \begin{figure}[htbp]
        \begin{minipage}[c]{.48\textwidth}
          \includegraphics[scale=0.23]{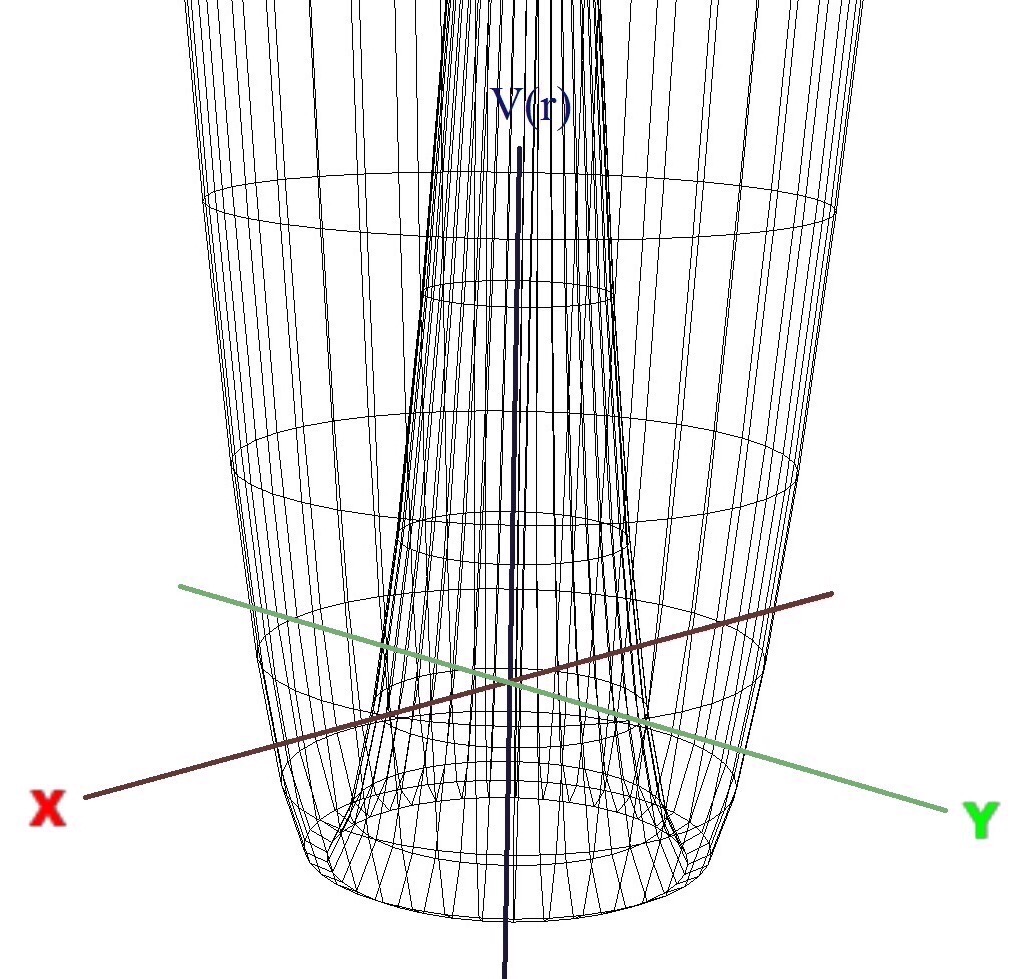}
          \caption{Three-dimensional plot of $V(r)$}
        \end{minipage}%
        \hspace{5mm}%
        \begin{minipage}[c]{.48\textwidth}
\begin{center}          \includegraphics[scale=0.5]{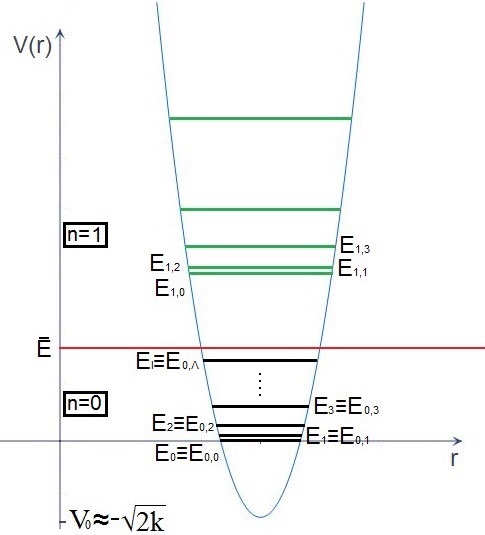}
          \caption{Two-dimensional plot of $V(r)$ including the energy-cutoff}
\end{center}        \end{minipage}
      \end{figure}

$\A_{\Lambda}:=P_{\Lambda}\Ob P_{\Lambda}$ and $\Hi_{\overline{E}}\equiv\Hi_{\Lambda}$ inherit the grading from $\Ob,\Hi$. For any $A^h\in\Ob^h$, $ |m|\leq{{\Lambda}}$,
\bea
\overline{A^h}\psi_m=\sum_{m'=-\Lambda}^{\Lambda}\psi_{m'} \langle\psi_{m'},A^h\psi_m\rangle=
\left\{\ba{ll}
 \psi_{m+h}\, \langle\psi_{m+h}, A^h\psi_m\rangle\quad &
\mbox{ if }\: |m|,|m+h|\leq{{\Lambda}},  \\[8pt]
0 & \mbox{otherwise.}
\ea\right.
\nonumber
\eea
In formula (\ref{flemma}) in the appendix  we compute the matrix element $\langle\psi_{m+h}, A^h\psi_m\rangle$ for 
$A$ of the form $A=f(\rho)e^{ih\varphi}=f(\rho)u^h$ [if $A$ contains also derivatives $\partial_\pm$ the
result can be expressed as a combination of matrix elements of the same type because of (\ref{mpartial})].
In particular, up to terms $O(1/k^{3/2})$
\be
\overline u\psi_m=\left\{\!\! \ba{l}
 \psi_{m+1}\,  \\[8pt]
0 \ea\!\!\!\!\right.,\quad
\overline{x}^+\psi_m=
\left\{\!\!\ba{ll}
 \psi_{m+1}\, \frac{a}{\sqrt{2}}\sqrt{1\!+\!\frac{m(m\!+\!1)}{k}}\quad &
\mbox{ if }-{{\Lambda}}\leq m\leq{{\Lambda}}-1, \\[8pt]
0 & \mbox{ otherwise,}
\ea\right.
\label{barx+}
\ee
where \ $a:=1\!+\!\frac 94\frac{1}{\sqrt{2k}}\!+\!\frac{137}{64k}$. \
Clearly \ $\sqrt{2}\,\overline{x}^+=\overline{u}\, a\sqrt{1\!+\!\frac{\overline{L}(\overline{L}\!+\!1)}{k}}$. \
As $u^\dagger,x^-$ are the adjoints of $u,x^+$, so are  $\overline{u^\dagger},\overline x^-$ respectively the adjoints of $\overline{u},\overline x^+$. We can get rid of the $m$-independent factor $a$ reabsorbing it in the redefinitions 
$$
\xi^\pm=\frac{\overline{x^\pm}}a.
$$
Therefore we find 
\bea
\overline{L}\psi_m=m\psi_m,\qquad\overline{H}=\overline{L}^2\!,\qquad  \xi^{\pm}\psi_m=
\left\{\!\!\ba{ll}
  \frac{1}{\!\sqrt{2}}\sqrt{1\!+\!\frac{m(m \pm 1)}{k}}\psi_{m\pm1} \:\: &
\mbox{if }-\!{{\Lambda}}\leq \pm m\leq{{\Lambda}}\!-\!1 \\[10pt]
0 & \mbox{otherwise,}
\ea\right.         \label{defLxiD=2}
\eea
[the second, third relations hold up to  terms $O(1/k^{1/2}), O(1/k^{3/2})$, respectively].
Eq. (\ref{defLxiD=2}) implies at leading order 
\bea
[\xi^+,\xi^-]\psi_m &=& \left\{\ba{ll} \displaystyle -\frac{m}{k}
\, \psi_m  \qquad &\mbox{if }\: | m|\leq{{\Lambda}}-1, \\[8pt]
\displaystyle \pm\frac{1}{2} \left[1+\frac{{{\Lambda}}({{\Lambda}}\!-\!1)}{k}\right]\, \psi_m  \qquad &\mbox{if }\: m=\pm{{\Lambda}} , \\[12pt]
0 & \mbox{otherwise.}
\ea\right. \label{crx+x-}\\[12pt]
\mathcal{R}^2\psi_m&=&  \left\{\ba{ll}
\displaystyle\left(1+\frac{m^2}{k}\right)\, \psi_m  \qquad &\mbox{if }\: | m|\leq{{\Lambda}}-1, \\[8pt]
\displaystyle \frac 12\left[1+\frac{{{\Lambda}}({{\Lambda}}-1)}{k}\right]\, \psi_m  \qquad &\mbox{if }\: m=\pm{{\Lambda}},\\[8pt]
0 & \mbox{otherwise;}
\ea\right.  \label{r^2D=2}
\eea
in (\ref{r^2D=2}) we have introduced  the square distance from the origin  $\mathcal{R}^2:={\xi}^+\xi^- +\xi^-\xi^+$,
in analogy with the classical definition.
We see that $\mathcal{R}^2$ is not 
identically equal to 1 on all of $\Hi_{{\Lambda}}$ (as in the 
standard quantization on the unit circle), but the $\psi_m$ 
are eigenvectors of $\mathcal{R}^2$ with eigenvalues depending only on $m^2$ and growing with $m^2$
(with the exception of the
states $\psi_{{\Lambda}},\psi_{{-\Lambda}}$ closest to the energy cutoff, which however play no role at lower energies); 
 physically this is to be expected because higher square angular momentum $m^2$ 
is equivalent to a larger centrifugal force, which classically yields a slightly more external circular trajectory.
Moreover, all these eigenvalues are  close to 1 and go  to 1 as $k\to\infty$, 
while the eigenfunctions factorize as $\psi_m(r,\varphi)\to \delta(r\!-\!1)e^{im\varphi}$.

If we now adopt (\ref{defLxiD=2}) as exact {\it definitions} of  $\overline{L}, \overline{H},\xi^+,\xi^-$, \ 
then (\ref{crx+x-}-\ref{r^2D=2}) are exact as well, and 
we easily find

\begin{propo}\label{2Dpropo}
The $\xi^+,\xi^-,\overline{L}$ defined by (\ref{defLxiD=2}) generate the $*$-algebra \ $\A_{\Lambda}=End(\mathcal{H}_{\Lambda})\simeq M_{2\Lambda+1}(\CC)$ \ of observables  on  $\mathcal{H}_{\Lambda}$; they
 fulfill \ $(\xi^+)^{2{{\Lambda}}+1}=(\xi^-)^{2{{\Lambda}}+1}=0$, \
$\prod\limits_{m=-{{\Lambda}}}^{{{\Lambda}}}\!\!(\overline{L}\!-\!mI)=0$,
\bea
\ba{l}
\xi^+{}^\dagger=\xi^-,\quad \overline{L}^\dagger=\overline{L},\qquad
\displaystyle \left[\overline{L},\xi^{\pm}\right]=\pm \xi^\pm, \qquad\left[\xi^+,\xi^-\right]=-\frac{\overline{L}}k+\mu\,\frac{\widetilde P_{{{\Lambda}}}\!-\!\widetilde P_{-{{\Lambda}}}}2,
\ea \label{comrelD=2}
\eea
where  $\widetilde P_{m}$ stands for the projector on the 1-dimensional subspace spanned by $\psi_m$ 
(clearly $P_{{{\Lambda}}}=\sum_{m=-\Lambda}^\Lambda\widetilde P_{m}$) and
$\overline{L}:=P_{{{\Lambda}}}LP_{{{\Lambda}}}$ is the projection on ${\cal H}_{{{\Lambda}}}$
of the angular momentum operator. \ All
$\widetilde P_m$  can be expressed
as polynomials in $\overline{L}$ using  the spectral decomposition of $\overline{L}$. 
Moreover, the square distance from the origin can be expressed as the  function of $\overline{L}^2$
\be
\mathcal{R}^2:={\xi}^+\xi^- +\xi^-\xi^+=  1+\frac{\overline{L}^2}{k} -
\mu\,\frac{\widetilde P_{{{\Lambda}}}\!+\!\widetilde P_{-{{\Lambda}}}}2, \qquad \mu:=1\!+\!\frac {{{\Lambda}}({{\Lambda}}\!+\!1)}{k}.        \label{R^2D=2}
\ee
\end{propo}
Of course, relations (\ref{comrelD=2})$_4$, (\ref{R^2D=2}) only hold at leading order in $1/\sqrt{k}$ if also (\ref{defLxiD=2}) do.

To obtain a fuzzy space depending only on one integer ${{\Lambda}}$ we can 
choose $k$ as a function of ${{\Lambda}}$ fulfilling (\ref{consistency}); the commutative
limit will be simply ${{\Lambda}}\to \infty$ (which implies $k\to \infty$).
One possible choice is 
\be
k={{\Lambda}}^2({{\Lambda}}\!+\!1)^2;\mbox{ then } (\ref{comrelD=2})_4
\mbox{ becomes }
[\xi^+,\xi^-]=\frac{-\overline{L}}{ {{\Lambda}}^2({{\Lambda}}\!+\!1)^2}+
\left[1\!+\!\frac 1{ {{\Lambda}}({{\Lambda}}\!+\!1)}\right]\!
\frac{\widetilde P_{{{\Lambda}}}\!-\!\widetilde P_{-{{\Lambda}}}}2.           \label{comrelD=2'}
\ee
Summarizing our results so far, we see that the combined effect of the confining potential and of the energy cutoff
is a non-vanishing commutator between the coordinates. Note that relations (\ref{comrelD=2}), (\ref{comrelD=2'}) are invariant not only under rotations
(this includes parity $\xi^a\mapsto -\xi^a$, $\overline{L}\mapsto \overline{L}$ in $D=2$), but also under 
orthogonal transformations with determinant $-1$, e.g. $\xi^1\mapsto \xi^1$,
$\xi^2\mapsto -\xi^2$, $\overline{L}\mapsto -\overline{L}$, i.e. under the whole group $O(2)$, as in the ordinary theory without cutoff.
This had to be expected, because both the commutation relations in the original infinite-dimensional model
and the Hamiltonian $H$ (hence also the projectors $P_{\overline E}$) are $O(2)$-invariant.
Apart from the sign and from the last term containing the projections, which plays no role far from the cutoff
${{\Lambda}}$, relations (\ref{comrelD=2}), (\ref{comrelD=2'}) are of  the Snyder's Lie algebra
type \cite{Snyder}, because the commutator of the coordinates is a generator of rotations.
 $\xi^+\!,\xi^-$ (or equivalently $\overline{x}^+\!,\overline{x}^-$) generate the whole $*$-algebra $\A_{\Lambda}$
(also $\overline{L}$ can be expressed as a non-ordered polynomial in $\xi^+\!,\xi^-$).

To compute $\overline{\partial_\pm}$ it is convenient to use polar coordinates. We find
\bea
\partial_\pm=\frac{1}{2x^{\pm}}\left(\partial_\rho\mp i\partial_\varphi\right),\qquad
\partial_\rho\psi_m=-(\rho\!-\!\widetilde{\rho}_m)\sqrt{k_m}\psi_m\propto \psi_{n=1,m}
\label{mpartial}
\eea
and $\overline{\partial_\rho}=0$, because of $P_{\Lambda}\psi_{n=1,m}=0$. Similarly,
$\overline{\partial_r}$, $\overline{\frac 1{x^\pm}\partial_\rho}$
go to zero in norm as $k\to\infty$. As consequences, neither $\partial_\pm\!-\!\overline{\partial_\pm}$
nor the commutator $\left[\overline{\partial_+},\overline{\partial_-}\right]$ vanish  as $k\to \infty$,
as expected. Only the vector field $L$ tangent to the circle survives with the correct
classical limit, as desired.  For completeness,
the actions of $\overline{\partial_+},\overline{\partial_-}$ are reported in (\ref{part+}-\ref{derfine}).

\subsection{Realization of the algebra of observables through $Uso(3)$}
\label{realD2}

For every $n\in\mathbb{N}$ the $*$-algebra $M_n(\mathbb{C})$ of endormophisms of $\mathbb{C}^n$ 
can be realized as the
$n=(2{{\Lambda}}\!+\!1)$-dimensional unitary representation $\pi_{{{\Lambda}}}$ of $so(3)\simeq su(2)$.
In fact, the operators on  $\mathcal{\Hi}_{{{\Lambda}}}$, and in particular $\overline u, \xi^\pm$,
are  naturally realized in $\pi_{{{\Lambda}}}\left[U\!so(3)\right]$, identifying $\psi_m$ as the
vectors $|m\rangle$ of the canonical basis, in standard ket notation. We denote as $E^+,E^-,E^0$
the Cartan-Weyl basis of $so(3)$,
\be
[E^+,E^-]=E^0,\qquad [E^0,E^\pm]=\pm E^\pm,\qquad E^\pm{}^\dagger=E^\mp,\qquad  E^0{}^\dagger=E^0,
 \label{su2rel}
\ee
as $\dagger$ (with an abuse of notation  also) its real structure, and as $C$ the  Casimir,
\bea
C=E^aE^{-a}=2E^+E^-\!+\!E^0(E^0\!-\!1)=2E^-E^+\!+\!E^0(E^0\!+\!1).                   \label{casimirsu2}
\eea 
The representation  $\pi_{{{\Lambda}}}$ is characterized by the Casimir eigenvalue 
$\pi_{{{\Lambda}}}(C)={{\Lambda}}({{\Lambda}}+1)$. We identify $\overline{L}=\pi_{{{\Lambda}}}(E^0)$, which will still determine the grading.
To simplify the notation in the sequel we drop $\pi_{{{\Lambda}}}$.
For every $A^h\in\Ob^h$ one can determine a function of one  variable $f_A(s)$ such that
$$
 \overline{ A^h}=f_A(E^0)E^h, \qquad \mbox{where } E^h=\left\{\ba{ll} (E^+)^h \quad &\mbox{if }\: h>0, \\
1 \quad &\mbox{if }\: h=0, \\
(E^-)^{-h} \quad &\mbox{if }\: h<0,
\ea\right.
$$
by requiring that $\langle\psi_{m+h}, A^h\psi_m\rangle=\langle\psi_{m+h}, f_A(E^0)E^h\psi_m\rangle$
and using (\ref{flemma}).
In particular  it is an easy exercise to check that from  (\ref{barx+}) and the adjoint relations  it follows 
\bea
\overline{L}=E^0, \qquad\qquad \xi^\pm=f_{\pm}(E^0)E^\pm
\label{transfD2}\
\eea 
and $\overline{u}=f_u(E^0)E^+$, \ where\footnote{Note that $f_u(E^0),f_{+}(E^0)$ 
are singular on $\psi_{{-\Lambda}}$, while $f_{x^-}(E^0)$ is  singular  
on $\psi_{{\Lambda}}$, but since their action follows that of $E^+$ or $E^-$ they can never act on such vectors, and the products  
at the right-hand side of (\ref{transfD2}) are well-defined on all of $\Hi_\Lambda$.
} 
$$
f_u(s)=\frac 1{\sqrt{{{\Lambda}}({{\Lambda}}+1)\!-\!s(s\!-\!1)}}, \qquad
f_{+}(s)=\frac 1{\sqrt 2}\,\sqrt{\frac{1\!+\!s(s\!-\!1)/k}{{{\Lambda}}({{\Lambda}}+1)\!-\!s(s\!-\!1)}}= f_{-}(s\!-\!1)=f_{-}(-s).      
$$
Therefore (\ref{transfD2}) fulfill (\ref{comrelD=2}). 
The inverse of the change of generators (\ref{transfD2}) is clearly
$$
E^0=\overline{L}, \qquad\qquad E^\pm=\left[f_{\pm}\left(\overline{L}\right)\right]^{-1}\, \xi^\pm.
$$
The eigenvalue condition $C={{\Lambda}}({{\Lambda}}+1)$  can be put more explicitly in either form
\bea
2E^-E^+={{\Lambda}}({{\Lambda}}+1)\!-\!E^0(E^0\!+\!1)=2\xi^-\xi^+\left[f_{+}\left(\overline{L}\!+\!1\right)\right]^{-2}.         \label{casimirsu2'}
\eea 
Summarizing, we have almost completely shown

\begin{propo}
Formulas (\ref{transfD2}) provide a $O(2)$-equivariant $*$-algebra  isomorphism between  the algebra $\A_{\Lambda}=End(\Hi_{{{\Lambda}}})$  of observables (endomorphisms) on 
$\Hi_{{{\Lambda}}}$  and that on the $C={{\Lambda}}({{\Lambda}}+1)$ irreducible representation of $Uso(3)$:
 \be
\A_{\Lambda}:=End(\Hi_{{{\Lambda}}})\simeq M_N(\CC)\simeq\pi_\Lambda[Uso(3)], \qquad N:=2\Lambda\!+\!1.                                 \label{isomD2}
\ee
\label{realso(3)'}
\end{propo}
(The $O(2)$-equivariance of this realization is shown below.)
Note also that every function of $E^0=\overline{L}$, including $f_{\pm}(E^0)$, can be expressed in polynomial form
by spectral decomposition.
As consequences, the generators $E^+,E^-$  of $U\!so(3)$ [which are characterized by
relations (\ref{su2rel}), where $E^0$ has to be understood as an abbreviation for the commutator of $E^+,E^-$]
further constrained by (\ref{casimirsu2'})$_1$, 
or alternatively $\xi^+,\xi^-$ fulfilling (\ref{comrelD=2})
 and (\ref{casimirsu2'})$_2$, generate all the algebra   $\A_{\Lambda} $; 
while ordered polynomials in $E^+,E^-,E^0$, or alternatively in $\xi^+,\xi^-,\overline{L}$, span $\A_{\Lambda} $.
Therefore the above results for the action of the operators $\xi^+,\xi^-,...$
on  ${\cal H}_{{{\Lambda}}}$ can be recovered determining the 
unique unitary representation of the $*$-algebra generated by $\xi^+,\xi^-$ fulfilling relations (\ref{comrelD=2}),
(\ref{casimirsu2'})$_2$, or more simply setting (\ref{transfD2}) and using our knowledge on the representation $\pi_\Lambda$
of $so(3)$. 

As known, the group of  $*$-automorphisms of $M_N(\CC)\simeq \A_{\Lambda}$ is inner and isomorphic
to $SU(N)$, i.e.
\be
a\mapsto g\, a \, g^{-1}, \qquad a\in \A_{\Lambda} ,   \label{autom}
\ee
with $g$ an unitary $N\times N$ matrix with unit determinant. A special role is played 
by the subgroup $SO(3)$ acting through the representation $\pi_\Lambda$, namely $g=\pi_\Lambda\left[e^{i\alpha}\right]$,
where $\alpha\in so(3)$, i.e. is a combination with real coefficients of $E^0, E^+\!+\!E^-,i(E^+\!-\!E^-)$.
In particular, choosing $\alpha=\theta E^0$ [i.e. in the adopted $so(2)$ Cartan subalgebra of $so(3)$] the automorphism amounts to a 
rotation in the $\overline x{}^1\overline x{}^2$ plane by an angle $\theta$, 
i.e. \ $E^0\mapsto E^0$ and $E^\pm\mapsto  e^{\pm i\theta} E^\pm$, \ or equivalently
\  $\overline{L}\mapsto \overline{L}$ \   and 
$$
\overline x^\pm\mapsto \overline x'{}^\pm = e^{\pm i\theta}\overline x^\pm \qquad
\Leftrightarrow\qquad\left\{ \ba{l } \overline x'{}^1=\overline x{}^1\cos\theta+\overline x{}^2\sin\theta,  \\ \overline x'{}^2=- \overline x{}^1\sin\theta +\overline x{}^2\cos\theta; \ea \right. 
$$ 
this a $SO(2)$ transformation in the $\overline x{}^1\overline x{}^2$ plane. Setting
$\alpha=\pi (E^++E^-)/\sqrt{2}$ we obtain a  $O(2)$ transformation
with determinant  $-1$ in such a plane; this amounts
to a rotation about $E^1:=(E^++E^-)/\sqrt{2}$ by an angle $\pi$, i.e. to  $E^0\mapsto - E^0$, $E^\pm\mapsto E^\mp$. As the functions $f_{\pm}$ 
fulfill $f_{\pm}(-s)=f_{\pm}(1\!+\!s)=f_{\mp}(s)$, this is equivalent to $\overline x{}^1\mapsto \overline x{}^1$,
$\overline x{}^2\mapsto -\overline x{}^2$, $\overline{L}\mapsto -\overline{L}$.
All other $O(2)$ transformations  
with determinant  $-1$ in the $\overline x{}^1\overline x{}^2$ plane can be obtained by composition with a $SO(2)$ transformation. $O(2)$ will play the role of isometry group of the fuzzy circle.


\subsection{Convergence to $O(2)$-equivariant quantum mechanics on $S$}

Here we explain in which sense our model converges to  $O(2)$-equivariant quantum mechanics on the circle
as $\Lambda\to\infty$.  

The $\psi_m\in\Hi_\Lambda$ are the fuzzy analogs of the $u^m$  considered just as elements  of an orthonormal
basis of the Hilbert space ${\cal L}^2(S)$.
Consider the $O(2)$-equivariant embedding ${\cal I}:\Hi_\Lambda\hookrightarrow {\cal L}^2(S)$ 
defined by 
$$
{\cal I}\left(\sum_{m=-\Lambda}^{\Lambda}\phi_m\psi_m\right)= \sum_{m=-\Lambda}^{\Lambda}\phi_mu^m.
$$
Below we shall drop the symbol ${\cal I}$ and simply identify $\psi_m=u^m$.
For all $\phi\in {\cal L}^2(S)$ let $\phi_ \Lambda:=\sum_{m=-\Lambda}^{\Lambda}\phi_mu^m$
(where $\{\phi_m\}_{m\in\ZZ}$ are the Fourier coefficients of $\phi$) be its projection on $\Hi_\Lambda$; clearly $\phi_ \Lambda\to\phi$ in the ${\cal L}^2(S)$-norm $\Vert\,\Vert$. 
In this sense  $\Hi_\Lambda$ invades ${\cal L}^2(S)$  as $\Lambda\to\infty$.

The embedding ${\cal I}$ induces the one 
${\cal J}\!:\!\A_\Lambda\!\hookrightarrow\! B\left[{\cal L}^2(S)\right]$; \ by definition, $\A_\Lambda$ annihilates
$\Hi_\Lambda^\perp$. 

The operators $L,\overline{L}$ coincide  on $\Hi_\Lambda$, and we easily check that 
 on the domain $D(L)\subset{\cal L}^2(S)$\footnote{$L$ is unbounded. \
$\phi\in D(L)$ amounts to \ $\sum\limits_{m\in\ZZ} m^2|\phi_m|^2<\infty$.} \ $\overline{L}\to L$ strongly as $\Lambda\to\infty$. Similarly, 
$f(\overline{L})\to f(L)$ strongly on $D[f(L)]$ for all measurable functions $f(s)$.

Bounded (in particular, continuous) functions $f$ on the circle, acting as multiplication operators 
$f\cdot:\phi\in{\cal L}^2(S)\mapsto f\phi\in{\cal L}^2(S)$, make up a subalgebra  $B(S)$ [resp. $C(S)$]
of $B\left[{\cal L}^2(S)\right]$. An element of $B(S)$ is actually an equivalence class $[f]$ of bounded functions differing
from $f$ only on a set of zero (Lebesgue-)measure, because for any $f_1,f_2\in[f]$ and $\phi\in{\cal L}^2(S)$
 $f_1\phi, f_2\phi$ differ only on a set of zero measure, and therefore are two equivalent representatives
of the same element of ${\cal L}^2(S)$.
Since $f$ belongs also to  ${\cal L}^2(S)$, by Carleson's theorem \cite{Car66}
 $f_N(\varphi):=\sum_{m=-N}^Nf_m e^{im\varphi}$ converges to $f(\varphi)$ as $N\to \infty$ for almost all $\varphi$, implying that 
$f_N(\varphi)\phi(\varphi)\to f(\varphi)\phi(\varphi)$ almost everywhere; in other words
 $f_\infty\in[f]$, where we have abbreviated
\be
f_\infty(\varphi):=\lim\limits_{N\to \infty}f_N(\varphi)=\lim\limits_{N\to \infty}
\sum_{m=-N}^Nf_m e^{im\varphi},                     \label{PlancherelFourier}
\ee
and each class can be identified by the corresponding Plancherel-Fourier series (\ref{PlancherelFourier}). 

The  natural fuzzy analog  of the vector space $B(S)$ is the vector space
of polynomials in $\xi^+$ (or $\xi^-$) of degree $2\Lambda$ at most, or equivalently
\be
{\cal C}_\Lambda:=\left\{\sum_{h=-2\Lambda}^{2\Lambda}f_h \eta^h\:,\: f_h\in\CC\right\}
\subset\A_\Lambda\subset B[{\cal L}^2(S)],
\label{def_CLambda}
\ee
where we have abbreviated \ $\eta^\pm:=\sqrt{2}\xi^\pm$ (so that $\eta^\pm\to u^{\pm 1}$),
$\eta^h:=(\eta^+)^h$ \ if $h\ge0$, \ $\eta^h:=(\eta^-)^{|h|}$ \ if $h<0$.  
In other words $\eta^h$ are the fuzzy analogs of the $u^h$ considered  as 
operators acting by multiplication on $\phi\in{\cal L}^2(S)$.

The operators $\eta^\pm$ converge {\it strongly} to $u^{\pm 1}$, because
\bea
(\eta^+\!-\!u)\phi=(\eta^+\!\!-\!u)\sum_{m\in\ZZ} \phi_mu^m=\!\sum_{m=-\Lambda}^{\Lambda-1}\!\!\left[\sqrt{\!1\!+\!\frac{m(m\!+\!1)}{k}}-1\right]\phi_m u^{m+ 1}-\!\!\!\!\!\!\sum_{m<-\Lambda, m\ge\Lambda}\!\!\!\!\!\!\!\!\phi_m u^{m+ 1}\quad \Rightarrow\nn
\Vert(\eta^+\!\!-\!u)\phi\Vert^2\le\sum_{m=-\Lambda}^{\Lambda-1}
\frac{m^2(m\!+\!1)^2}{4k^2}|\phi_m|^2+\!\sum_{| m|\ge\Lambda}\!\!|\phi_m|^2
\le\frac{\Lambda^2(\Lambda\!+\!1)^2}{4k^2}\Vert \phi\Vert^2+\!\sum_{| m|\ge\Lambda}|\phi_m|^2, \qquad\label{ineqeta}
\eea
and by  (\ref{consistency})  the rhs goes to zero as $\Lambda\to\infty$; the first inequality follows from 
$$
|m|\!\le\!\Lambda\quad\Rightarrow\quad 0\!\le\! m(m\!+\!1) \!\le\! \Lambda(\Lambda\!+\!1), \qquad
\varepsilon\!>\!0\quad\Rightarrow\quad 
\sqrt{1\!+\!\varepsilon}-1\!<\! \varepsilon/2.
$$ 
Similarly one shows that $\eta^-\to u^{-1}$. 
Since
for all $\Lambda>0$  $\eta^\pm$ vanish on $\Hi_\Lambda^\perp$,
then choosing $\phi=u^{\Lambda+1}$ one finds
$\eta^+\phi=0$, $\Vert(\eta^+\!-\!u)\phi\Vert=\Vert-u^{\Lambda+2}\Vert=1$, 
implying $\Vert\eta^+\!-\!u\Vert\ge 1$ for all $\Lambda$.
This  prevents 
$\eta^\pm$  to converge  to $u^{\pm 1}$ {\it in operator norm}.

The previous result extends to all $f\in B(S)$; in particular, to $f\in C(S)$. Let
\be
\hat f_\Lambda:=\sum_{h=-2\Lambda}^{2\Lambda}f_h \eta^h\in\A_\Lambda\subset B[{\cal L}^2(S)]
\label{def_hatfLambda}.
\ee

\begin{propo}\label{2Dconverge}
If we choose $k(\Lambda)\ge 2 {{\Lambda}}({{\Lambda}}\!+\!1)(2{{\Lambda}}\!+\!1)^2$, then 
for all $f,g\in B(S)$ the following strong limits as $\Lambda\to\infty$ hold:   $\hat f_\Lambda\to f\cdot$,  $\widehat{(fg)}_\Lambda\to fg\cdot$, and $\hat f_\Lambda\hat g_\Lambda\to fg\cdot$.
\end{propo}
\noindent
(the proof is in  appendix \ref{2Dconvergence}).
The last statement says that  the product in $\A_\Lambda$ of the approximations $\hat f_\Lambda,\hat g_\Lambda$ goes to the product in $ B\left[{\cal L}^2(S)\right]$ of $f\cdot,g\cdot$.

\section{$O(3)$-equivariant fuzzy sphere}
\label{D=3}

When $D=3$  one can associate a pseudovector $L_i=\frac{1}{2}\varepsilon^{ijk}L_{jk}$ to the antisymmetric matrix $L_{ij}$ of the angular momentum components.
For all vectors $\bv$ depending on $\bx,i\nabla$
we shall use either the components $v^i$ or the ones $v^a$, $a\in\{-,0,+\}$, defined by
\be
\left(\begin{array}{c} v^+\\ v^-\\ v^0\\
\end{array}\right)  = \: 
\underbrace{\!\!\left(
\begin{array}{ccc}
\frac{1}{\sqrt{2}}&\frac{i}{\sqrt{2}}&0\\
\frac{1}{\sqrt{2}}&\frac{-i}{\sqrt{2}}&0\\
0&0&1\\
\end{array}
\right)\!\!}_{U} \left(\begin{array}{c} v^1\\ v^2\\ v^3\\
\end{array}\right).\label{trasfcoord}
\ee
($U$ is a unitary matrix) which fulfill  
\bea
 [L_a,v^a]=0, \qquad [L_0,v^\pm]=\pm v^\pm, \qquad [L_\pm,v^\mp]=\pm  v^0, \qquad
[L_\pm,v^0]=\mp  v^\pm.          \label{basic_so3'}
\eea
In particular, \ $x^0\!\equiv\!z$, $x^\pm\!=\!\frac{x^1\pm ix^2}{\sqrt{2}}\!\equiv\!\frac{x\pm iy}{\sqrt{2}}\!=\!\frac{r\sin{\theta}e^{\pm i\varphi}}{\sqrt{2}}$. 
\ We set $t^a\!:=\!\frac{x^a}{r}$. Correspondingly, the metric matrix \ $\eta_{ij}=\delta_{ij}$  \ becomes \
$\widetilde{\eta}_{ab}=(U\eta U^T)_{ab}=\delta_{-ab}$. \
Moreover,  (\ref{LaplacianD}) becomes \ $\Delta=\frac 1r\partial_r^2 r-\frac 1{r^2}L^2$. \
We look for $\psi$ of the form $\psi_l^m\!\left(r,\theta,\varphi\right)=\frac{f(r)}{r}Y_l^m\!\left(\theta,\varphi\right)$,
 where  $Y_l^m\!\left(\theta,\varphi\right)$ are the spherical harmonics:
$$
L^2\, Y_l^m\!\left(\theta,\varphi\right)=l(l+1)Y_l^m\!\left(\theta,\varphi\right),\hspace{1cm}\hspace{1cm}L_3\,Y_l^m\!\left(\theta,\varphi\right)=mY_l^m\!\left(\theta,\varphi\right).
$$
Equation (\ref{eqpolar}) becomes \ $-f''(r)+\left[V(r)+\frac{l(l+1)}{r^2}-E\right]f(r)=0$. \
By condition (\ref{cond1}),  in the region $|r\!-\!1|\le \sqrt{\frac{\overline{E}\!-\!V_0}{2k}}$
we can neglect the terms of order higher than two
in the Taylor expansion of \  $\frac{1}{r^2}=1-2(r-1)+3(r-1)^2+...$, \ $V(r)=V(1)+2k(r-1)^2+...$ \ and
thus  approximate this equation  by the eigenvalue equation for a $1-$dimensional harmonic oscillator, that is
\bea
&&-f''(r)+k_l(r-\widetilde{r}_l)^2f(r)=\widetilde{E} f(r)\qquad\label{harmoscD=3}\\[10pt]
&&\mbox{with }\qquad k_l:=2k+3l(l+1)\: ,\nn
&&\qquad\qquad\widetilde{r}_l:=\frac{2k+4l(l+1)}{2k+3l(l+1)}=1+\frac{l(l+1)}{2k}-\frac{3l^2\left(l+1\right)^2}{4k^2}+O\left(k^{-3} \right),\nn[8pt]
&&\qquad\qquad\widetilde{E}:=E-V(1)-l(l+1)+\frac{l^2(l+1)^2}{2k+3l(l+1)}\: .
\nonumber
\eea
The square-integrable solutions of (\ref{harmoscD=3}) have the form
$$
f_{n,l}(r)=N_l e^{-\frac{\left(r-\widetilde{r}_l\right)^2\sqrt{k_l}}{2}}H_n\left(\left(r-\widetilde{r}_l\right)\sqrt[4]{k_l}\right),
\qquad n=0,1,....,
$$
and $\widetilde{E}=(2n+1)\sqrt{k_l}$.
Choosing $V(1)$ such that in the lowest level (characterized by $n=l=0$) $E=0$, one has \
$V(1)=-\sqrt{k_0}=-\sqrt{2k}$ and, consequently,
$$
E\equiv E_{n,l}:=l(l+1)+(2n+1)\sqrt{2k\!+\!3l(l\!+\!1)}+V(1)-\frac{l^2(l+1)^2}{2k\!+\!3l(l\!+\!1)}= l(l+1)+2n\sqrt{2k}+O\left(\frac{1}{\sqrt{k}}\right)
$$
The term $l(l+1)$ gives exactly what we wish, (part of) the spectrum of $L^2$ (the Laplacian on the sphere).
To eliminate the subsequent, undesidered term  we fix the energy cutoff
$\overline E<2\sqrt{2k}$ so as to exclude all the states with $n>0$, i.e.  ``freeze'' radial oscillations: $n=0$. 
Therefore we impose an energy cut-off $\overline{E}=\Lambda(\Lambda+1)$, that is we project the theory to the finite-dimensional subspace 
 $\Hi_{\overline{E}}\equiv\mathcal{H}_{\Lambda}\subset \Hi$ 
 spanned by the $\psi_l^m:=\psi_{0,l,m}$ with $|m|\leq l$ and $l\leq\Lambda$. We 
 denote as  $P_{\Lambda}$ the projection over $\mathcal{H}_{\Lambda}$ and abbreviate $E_{l}=E_{0,l}$.
For consistency one must choose
\be
\Lambda(\Lambda+1)\leq2\sqrt{2k}. \label{consistency3D}
\ee
The spectrum of $\overline{H}$ becomes the whole spectrum $\{l(l+1)\}_{l\in \NN_0}$ of   $L^2$ 
in the limit $\Lambda,k \to\infty$ respecting (\ref{consistency3D}).
The eigenfunction of $L^2,L_3,H$ with eigenvalues respectively $l(l+1)$, $m$, $E_l=l(l\!+\!1)\!+\!O\left(\frac{1}{\sqrt{k}}\right)$ is
\be
\psi_l^m(r,\theta,\varphi)= \frac{N_l}{r}e^{-\frac{\left(r-\widetilde{r}_l\right)^2\sqrt{k_l}}{2}}\, Y_l^m(\theta,\varphi)
\ee
at the leading order in $k$. The actions of the  $\overline{L}_a$  ($a\in\{-1,0,+1\}$) and $\overline{H}$ are  therefore
\be
\overline{L}_0\psi_l^m=m\,\psi_l^m,\quad
\overline{L}_{\pm}\psi_l^m=\frac{\sqrt{(l\!\mp\! m)(l\!\pm\! m\!+\!1)}}{\sqrt{2}}\psi_l^{m\pm 1}\!=:\!\gamma_{l}^{\pm,m}\psi_l^{m\pm 1},\quad \overline{H}\psi_l^m=l(l\!+\!1)\,\psi_l^m
\label{defLD=3}
\ee
[the last relation holds up to \ $O(1/\sqrt{k})$].
In the appendix we compute the  normalization factor $N_l$;  moreover,
we show that the 
action of $\overline{x}^a$ on the vectors $\psi_{l}^m$ reads
\bea
\overline{x}^a\psi_{l}^m=\left\{\!\!
\ba{ll}
c_l A_{l}^{a,m}\psi_{l-1}^{m+a}+
c_{l+1} B_{l}^{a,m} \psi_{l+1}^{m+a
}&\mbox{ if }l<\Lambda,\\[8pt]
c_lA_l^{a,m}\psi_{\Lambda-1}^{m+a}&\mbox{ if }
l=\Lambda,\\[8pt]
0&\mbox{otherwise,}
\ea
\right.   \label{barxpsiD3}
\eea
where $A^{a,m}_l,B^{a,m}_l$ are the coefficients involved in the formula
\be
\label{tY'}
t^{a} Y^m_l=A^{a,m}_l Y^{m+a}_{l-1}+B^{a,m}_lY^{m+a}_{l+1},                           
\ee
which are explicitly reported in (\ref{ClebschAB}), while, up to terms $O(1/k^{3/2})$,
\bea
c_l= \sqrt{1+\frac{l^2}{k}}\qquad 1\le l\le \Lambda,\qquad c_0=c_{\Lambda+1}=0. \label{defc_l}
\eea
We now adopt (\ref{defLD=3}-\ref{defc_l}) as exact {\it definitions} of  $\overline{x}^a,\overline{L}^a,\overline{H}$.
The $\overline{x}^i,\overline{L}^i$ can be obtained by the inverse transformation of (\ref{trasfcoord}).
In the appendix we prove

\begin{propo}\label{3Dpropo}
The $\overline{x}^i,\overline{L}_i$ defined by (\ref{defLD=3}-\ref{defc_l}), (\ref{trasfcoord}) generate the $*$-algebra 
\ $\A_{\Lambda}:=End(\mathcal{H}_{\Lambda})\simeq M_{(\Lambda+1)^2}(\CC)$ \ of observables  on  $\mathcal{H}_{\Lambda}$. They
 fulfill 
\bea
&& \prod_{l=0}^{\Lambda}\left[\overline{L}^2-l(l+1)I\right] =0,\qquad
\prod_{m=-l}^{l}{\left(\overline{L}_3-mI\right)}\widetilde{P}_l=0,\qquad \left(\overline{x}^{\pm}\right)^{2\Lambda+1}=0, \label{rf3D3}\\[4pt]
&& \overline{x}^{i\dag}=\overline{x}^i, \qquad 
\overline{L}_i ^{\dag}=\overline{L}_i, \qquad [\overline{L}_i,\overline{x}^j]=i\varepsilon^{ijh}\overline{x}^h, \qquad 
\left[\,\overline{L}_i,\overline{L}_j\right]=i\varepsilon^{ijh}\overline{L}_h, \label{rf3D4}\\[8pt]
&& \overline{x}^i\overline{L}_i=0,\hspace{1.5cm}
[\overline{x}^i,\overline{x}^j]=i\varepsilon^{ijh}\left(-\frac{I}{k}+K\widetilde{P}_{\Lambda}\right)\overline{L}_h \hspace{1.5cm}i,j,h\in\{1,2,3\}\label{xx}
\eea
where $\overline{L}^2:=\overline{L}_i\overline{L}_i=\overline{L}_a\overline{L}_{-a}$ is  $L^2$ projected on $\Hi_\Lambda$,
$\widetilde{P}_l$ is the projection on its eigenspace  with eigenvalue
$l(l+1)$, and $K=\frac{1}{k}+\frac{1+\frac{\Lambda^2}{k}}{2\Lambda+1}$. Moreover, the square distance from the origin is
\be
\mathcal{R}^2:=\overline{x}^{i}\overline{x}^{i}=\overline{x}^{a}\overline{x}^{-a}= 1
+\frac{\overline{L}^2+1}{k}-\left[1+\frac{(\Lambda\!+\!1)^2}{k}\right]\frac{\Lambda\!+\!1}{2\Lambda+1}\widetilde P_{\Lambda}.
\label{R^2D=3}
\ee
\end{propo}
By the last equation, again $\mathcal{R}^2$ can be expressed as a function of $\overline{L}^2$ only, grows with the latter, 
and its spectrum collapses to 1 (apart from the highest eigenvalue) as $k\to\infty$. 

Of course, relations (\ref{xx})$_4$, (\ref{R^2D=3}) hold only at leading order in $1/\sqrt{k}$ if also (\ref{defLD=3}-\ref{defc_l}) do.

To obtain a fuzzy space depending only on one integer $\Lambda$ we can choose $k$ as a function of $\Lambda$ fulfilling (\ref{consistency3D}); the commutative limit will be simply $\Lambda\rightarrow+\infty$ (what implies $k\rightarrow+\infty$). One possible choice is $k=\Lambda^2(\Lambda+1)^2$; then (\ref{xx}) becomes 
\bea
\left[\,\overline{x}^i,\overline{x}^j\right]=i\varepsilon^{ijk}\left[-\frac{I}{\Lambda^2(\Lambda+1)^2}+\left(\frac{1}{\Lambda^2(\Lambda+1)^2}+\frac{1+\frac{1}{(\Lambda+1)^2}}{2\Lambda+1} \right)\widetilde{P}_{\Lambda}\right]\overline{L}_k\label{rf3D1}
\eea
and $\mathcal{R}^2\rightarrow 1$ as well.\\
We note that relations (\ref{rf3D4}), (\ref{xx}), (\ref{rf3D1}) are similar to those defining  the Snyder's Lie algebra, because the commutator of the coordinates is a polynomial in the generator of rotations $\overline{L}_i$, more precisely proportional to the $\overline{L}_i$ apart on $\Hi_\Lambda$, and therefore are invariant under parity $\overline{x}^a\rightarrow-\overline{x}^a$ (because $L_3$ and $L^2$ are), contrary to the fuzzy sphere of Madore.

The operators $\overline{\partial_a}=P_{\Lambda}\frac{\partial}{\partial x^a}P_{\Lambda}$ are such that
$$
\langle \psi_{l'}^{m'},\partial_a\psi_l^m\rangle\neq 0 \quad \Rightarrow \quad l'=l\pm1,\: m'=m-a;
$$
 the explicit actions of the $\overline{\partial}_a$ 
and of the commutator 
$\left[\overline{\partial}_a,\overline{\partial}_b\right]$ are  in the appendix. As with $D=2$, the action of the $\overline{\partial}_a$ on the eigenfunction $\psi_l^m$ gives a vector which has a non trivial projection on the Hilbert subspace corrisponding to $n=1$. Consequently, neither $\partial_a-\overline{\partial}_a$ nor the commutator $\left[\overline{\partial}_a,\overline{\partial_b}\right]$ vanish  as $k\rightarrow+\infty$, i.e. $\overline{\partial_a}$ has not the usual commutative limit.

\subsection{Realization of the algebra of observables through $Uso(4)$}
\label{realso(4)}

As the Lie algebra  $su(2)$ is spanned by $\left\{E_i\right\}_{i=1}^3$ fulfilling 
\be
\left[E_i,E_j\right]=i\varepsilon^{ijk}E_k,
\ee
 $so(4)\simeq su(2)\oplus su(2)$ \ is spanned by
$\left\{E_i^1,E_i^2\right\}_{i=1}^3$, where 
we have abbreviated \ $E_i^1:=E_i\otimes 1$, $E_i^2:= 1\otimes E_i$, and
\bea
[E^1_i, E^2_j]=0,\qquad [E^1_i, E^1_j]=i\varepsilon^{ijk}E^1_k,\qquad [E^2_i, E^2_j]=i\varepsilon^{ijk}E^2_k.
\label{CRE}
\eea
$L_i=E_i^1+E_i^2$ and  $X_i=E_i^1-E_i^2$ make up  an alternative basis  of $so(4)$ and fulfill
\bea
[L_i, L_j]=i\varepsilon^{ijk}L_k, \qquad[L_i, X_j]=i\varepsilon^{ijk}X_k,\qquad [X_i, X_j]=i\varepsilon^{ijk}L_k.
\label{CRXL}
\eea
The $L_i$ close another $su(2)$ Lie algebra. 
Applying the transformation (\ref{trasfcoord}) we obtain 
alternative generators labelled by $a\in\{-,0,+\}$, fulfilling
\bea
\left[L_+,L_-\right]=L_0,\qquad \left[L_0,L_{\pm}\right]=\pm L_{\pm}=[X_0,X_\pm],\qquad [X_+,X_-]= L_0, \label{u1}\\[6pt]
 [L_\pm,X_\mp ]=\pm X_0,\qquad [L_0,X_\pm]=\pm X_\pm= [X_0,L_\pm],
\qquad   [L_a,X_a]=0 \label{u2}
\eea
(no sum over $a$), where we have abbreviated \ $L^2=L_iL_i=L_aL_{-a},\quad X^2=X_iX_i=X_aX_{-a}$. 

Let $\pi_j$ be the unitary irreducible representation  of $Usu(2)$ on the $(2j\!+\!1)$-dimensional Hilbert space $V_j$; this is characterized by the 
eigenvalue $j(j\!+\!1)$ of the Casimir  $C:=E_iE_i$. 
The tensor product representation $\bpi_{\Lambda}:=\pi_{\frac{\Lambda}{2}}\otimes  \pi_{\frac{\Lambda}{2}}$\  of $Uso(4)\simeq U\!su(2)\otimes U\!su(2)$ \ on the Hilbert space 
${\bf V}_{\Lambda}:=V_{\frac{\Lambda}{2}}\otimes V_{\frac{\Lambda}{2}}$ is characterized by the conditions $C^1:=E^1_iE^1_i=\frac{\Lambda}{2}(\frac{\Lambda}{2}+1)=E^2_iE^2_i=:C^2$, or equivalently
\be
X\cdot L=L\cdot X=0,\qquad X^2\!+\!L^2=\Lambda(\Lambda\!+\!2)\label{u3};
\ee
here and below we  drop the symbol $\bpi_{\Lambda}$.  ${\bf V}_{\Lambda}$ 
admits an orthonormal basis consisting of common eigenvectors of $L^2$ and $L_0$:
\be
L_0\left\vert l,m\rangle\right.=m\left\vert l,m\rangle\right.\quad L^2\left\vert l,m\rangle\right.=l(l+1)\left\vert l,m\rangle\right.\quad \mbox{with } 0\leq l\leq \Lambda\mbox{ and }|m|\leq l ,
\ee
in standard ket notation. ${\bf V}_{\Lambda},\Hi_{\Lambda}$ have
the same dimension $(\Lambda\!+\!1)^2$ and the same decomposition in irreducible
representations of the $L_i$ subalgebra, and will be eventually identified.

We determine the action of the $X_a$ on the $\vert l,m\rangle$. Because of the commutation relations $[L_0,X_a]=aX_a$
it must be $X_a\left\vert l,m\rangle\right.=\sum_{j=0}^{\Lambda}\alpha_{l,j}^{a,m}\left\vert j,m+a\rangle\right. $.
In the appendix we show that $\alpha_{l,j}^{a,m}=0$ unless $j=l\pm 1$, and more precisely that the previous relations are fulfilled by
\be
X_a\left\vert l,m\rangle\right.=d_l A_l^{a,m}\left\vert l-1,m+a\rangle\right.+d_{l+1}B_l^{a,m}\left\vert l+1,m+a\rangle\right., \qquad
d_{l}:=\sqrt{(\Lambda\!+\!1)^2-l^2}.\label{u4}
\ee
The operators on $\mathcal{\Hi}_{{{\Lambda}}}$, and in particular $\overline{L_a}, \overline{x}^a$,
are naturally realized in $\bpi_{\Lambda}\left[U\!su(2)\otimes Usu(2)\right]$, identifying $\psi_l^m$ as the
vectors of the canonical basis $|l,m\rangle$. For simplicity,
we introduce the operator $\lambda:=[\sqrt{4L^2+1}-1]/2$; $\vert l,m\rangle$ is an eigenvector with eigenvalue $l$.
The Ansatz
\be
\overline{L}_a=L_a,\qquad \overline{x}^a=g^*(\lambda)\, X_a\,g(\lambda), \label{transfD3}
\ee
automatically fulfills the the hermiticity relations $\overline{x}^a{}^\dagger=\overline{x}^{-a}$. Applying  $\overline{x}^a$ to $\vert l,m\rangle$
we find
\be
\overline{x}^a\left\vert l,m\rangle\right.=g(l)g^*(l-1)d_l A_l^{a,m}\left\vert l-1,m+a\rangle\right.+g^*(l+1)g(l)d_{l+1}B_l^{a,m}\left\vert l+1,m+a\rangle\right. ;
\ee
this agrees with (\ref{barxpsiD3}) if and only if for $l>1$
\be
g^*(l-1)g(l)=\frac{c_l}{d_l}=\frac{\sqrt{1+\frac{l^2}{k}}}{\sqrt{(\Lambda\!+\!1\!-\!l)(\Lambda\!+\!1\!+\!l)}},     
\label{gcond} 
\ee
which is solved by
\bea
g(l) &=& \sqrt{\frac{\prod_{h=0}^{l-1}(\Lambda\!+\!l\!-\!2h)}{\prod_{h=0}^l(\Lambda\!+\!l\!+\!1\!-\!2h)}
\prod_{j=0}^{\left[\frac{l\!-\!1}2\right]}\frac{1+\frac{(l\!-\!2j)^2}k}{1+\frac{(l\!-\!1\!-\!2j)^2}k}}\quad,
\label{gO(3)'}
\eea
where again $[b]$ stands for the integer part of $b$. Alternatively,
using the  basic property  $\Gamma(z+1)=z \Gamma(z)$ of the  Euler gamma function we can express a solution  in the form
\be\label{gO(3)}
g(l)=\sqrt{
\frac{\Gamma\!\left(\frac {\Lambda\!+\!l}2\!+\!1\right)\Gamma\!\left(\frac {\Lambda\!-\!l\!+\!1}2\right)}
{\Gamma\!\left(\frac {\Lambda\!+\!1\!+\!l}2\!+\!1\right)\Gamma\!\left(\frac {\Lambda\!-\!l}2\!+\!1\right)}
\frac{\Gamma\!\left(\frac l2\!+\!1\!+\!\frac{i\sqrt{k}}2\right)\Gamma\!\left(\frac l2\!+\!1\!-\!\frac{i\sqrt{k}}2\right)}
{\sqrt{k}\:\Gamma\!\left(\frac {l\!+\!1}2\!+\!\frac{i\sqrt{k}}2\right)\Gamma\!\left(\frac {l\!+\!1}2\!-\!\frac{i\sqrt{k}}2\right)}}
\ee
(see the Appendix); this makes sense also for generic complex argument $l$.
The inverse of the transformation (\ref{transfD3}) is clearly $X_a=[g^*(\lambda)]^{-1}\,\overline{x}^a\,[g(\lambda)]^{-1}$.

We have thus proved by an explicit construction 

\begin{propo}
Formulas (\ref{transfD3}), (\ref{gO(3)}) define a $O(3)$-equivariant $*$-algebra  isomorphism between  the algebra $\A_{\Lambda}=End(\Hi_{{{\Lambda}}})$   of observables (endomorphisms) on 
$\Hi_{{{\Lambda}}}$  and the  $C_1=C_2=\frac{{\Lambda}}2\left(\frac{{\Lambda}}2+1\right)$ irreducible representation of  \ $Uso(4)\simeq U\!su(2)\otimes U\!su(2)$:
\be
\A_{\Lambda}:=End(\Hi_{{{\Lambda}}})\simeq M_N(\CC)\simeq\bpi_\Lambda[Uso(4)], \quad N:=(\Lambda\!+\!1)^2.             \label{isomD4}
\ee
\label{realso(4)'}
\end{propo}

\medskip
As already recalled, the group of  $*$-automorphisms of $M_N(\CC)\simeq \A_{\Lambda}$ is inner and isomorphic
to $SU(N)$, i.e. of the type (\ref{autom})
with $g$ an unitary $N\times N$ matrix with unit determinant. A special role is played 
by the subgroup $SO(4)$ acting in the representation $\bpi_\Lambda$, namely $g=\bpi_\Lambda\left[e^{i\alpha}\right]$,
where $\alpha\in so(4)$.
In particular, choosing $\alpha=\alpha_iL_i $ ($\alpha_i\in\RR$)  the automorphism amounts to 
  a $SO(3)\subset SO(4)$ transformation (a rotation in 3-dimensional space). Parity $(L_i,X_i)\mapsto (L_i,-X_i)$, or equivalently
$E_i^1\leftrightarrow E_i^2$ [the only automorphism
of $so(4)$ corresponding to the exchange of the two nodes in the Dynkin diagram], is a $O(3)\subset SO(4)$ transformation
with determinant  $-1$ in the $X_1X_2X_3$ space, and therefore also
in the $\bar x^1\bar x^2\bar x^3$ space. This shows that (\ref{transfD3}) is equivariant
under $O(3)$, which plays the role of isometry group of this fuzzy sphere.

\subsection{Convergence to $O(3)$-equivariant quantum mechanics on $S^2$}
\label{3Dconverge}

Here we explain in which sense our model converges to  $O(3)$-equivariant quantum mechanics on the sphere
as $\Lambda\to\infty$.  

The $\psi_l^m\in\Hi_\Lambda$ are the fuzzy analogs of the spherical
harmonics $Y_l^m$  considered just as elements  of an orthonormal
basis of the Hilbert space ${\cal L}^2(S)$. 
The decomposition
of $\Hi_\Lambda$ into irreducible components under $O(3)$ reads
\be
\qquad\qquad \Hi_\Lambda=\bigoplus\limits_{l=0}^{\Lambda} V_l,\qquad\qquad V_l:=
\left\{\sum_{m=- l}^l\phi^m \psi^m_l\:,\: \phi^m\in\CC\right\}. \label{deco1}
\ee
(\ref{deco1})$_1$ becomes the
decomposition of ${\cal L}(S^2)$ in the limit $\Lambda\to\infty$. 
Consider the $O(3)$-covariant embedding ${\cal I}:\Hi_\Lambda\hookrightarrow {\cal L}^2(S)$ 
defined by 
$$
{\cal I}\left(\sum_{l=0}^{\Lambda}{\sum_{m=-l}^{l}\phi_l^m\psi_l^m}\right)= \sum_{l=0}^{\Lambda}\sum_{m=-l}^{l}\phi_l^m Y_l^m
$$
Below we shall drop the symbol ${\cal I}$ and simply identify $\psi_l^m=Y_l^m$.
For all $\phi\in {\cal L}^2(S^2)$ let 
$\phi_ \Lambda:=\sum_{l=0}^{\Lambda}\sum_{m=-l}^{l}\phi_l^mY_l^m
$,
where $\{\phi_l^m\}_{l,m}$ are the coefficients of the decomposition of $\phi$ in the orthonormal basis of spherical harmonics; clearly $\phi_ \Lambda\to\phi$
in the ${\cal L}^2(S^2)$-norm $\Vert\,\Vert$. In this sense  $\Hi_\Lambda$ invades ${\cal L}^2(S^2)$
 as $\Lambda\to\infty$.

The embedding $\mathcal{I}$ induces the one $\mathcal{J}:\mathcal{A}_{\Lambda}\hookrightarrow B\left[\mathcal{L}^2\left(S^2\right)\right]$; by construction, $\mathcal{A}_{\Lambda}$ annihilates $\mathcal{H}_{\Lambda}^{\perp}$.

\medskip
The operators $L_i,\overline{L}_i$ coincide on $\Hi_\Lambda$, and we can easily check that on the domain $D\left(L_i\right)\subset \mathcal{L}^2\left(S^2\right)$\footnote{$L_i$ is unbounded, for example $\phi\in D(L_0)$ implies $\sum_{l\in\mathbb{N}_0}\sum_{|m|\leq l} m^2|\phi_l^m|^2<\infty$.} $\overline{L}_i\to L_i$
strongly as $\Lambda\to\infty$. Similarly, 
$f(\overline{L}_i)\to f(L_i)$ strongly on $D[f(L_i)]$ for all measurable function $f(s)$.

\medskip
Bounded (in particular, continuous) functions $f$ on the sphere, acting as multiplication operators 
$f\cdot:\phi\in{\cal L}^2(S^2)\mapsto f\phi\in{\cal L}^2(S^2)$, make up a subalgebra  $B(S^2)$  [resp. $C(S^2)$]
of $B\left[{\cal L}^2(S^2)\right]$. An element of $B(S^2)$ is actually an equivalence class $[f]$ of bounded functions differing
from $f$ only on a set of zero measure, because for any $f_1,f_2\in[f]$ and $\phi\in{\cal L}^2(S^2)$
 $f_1\phi, f_2\phi$ differ only on a set of zero measure, and therefore are two equivalent representatives of the same element of ${\cal L}^2(S^2)$.
Since $f$ belongs also to  ${\cal L}^2(S^2)$, then 
$f_N(\theta,\varphi):=\sum_{l=0}^N\sum_{m=-l}^nf_l^m Y_l^m(\theta,\varphi)$ converges to $f(\theta,\varphi)$ in the ${\cal L}^2(S^2)$ norm as $N\to \infty$.

To introduce the fuzzy analogs of  $B(S^2)$ and $f$ we need to introduce first the fuzzy analogs
$\widehat{Y}_l^m$ of the spherical harmonics  $Y_l^m$ seen as elements of $B(S^2)$ 
(acting by multiplication $Y_l^m\cdot$: $\psi\mapsto Y_l^m\psi$).
We recall that the $Y_l^m$ are
trace-free, homogenous polynomials of degree $l$ in the
$t^a$; as the $Y_l^m$ with fixed $l$  span  the $(2l+1)$-dim irreducible
representation $V_l$ of $so(3)$, they can be obtained recursively from the
highest weight $Y^l_l=M_l(t^+)^l$ 
(the absolute value of the normalization factor is $|M_l|=\sqrt{\frac{(2l+1)!!}{4\pi (2l)!!}}\le 1/\sqrt{4\pi}$)
applying powers of $L_-$, $\sqrt{2}L_-Y^m_l=\sqrt{(l\!+\! m)(l\!-\! m\!+\!1)}Y^{m-1}_l$, implying
$$
Y_l^m=M_l\sqrt{\frac{(l+m)!2^{l-m}}{(2l)!(l-m)!}} \:\:
L_-^{l-m}(t^+)^l.
$$
We therefore define the $\widehat{Y}_l^m$ by the formulae 
\be
\widehat{Y}_l^m:=M_l\sqrt{\frac{(l+m)!2^{l-m}}{(2l)!(l-m)!}} \:
L_-^{l-m}(\overline{x}^+)^l.                  \label{defhatY}
\ee
By use of $L_-\overline{x}^+=\overline{x}^0$, $L_-\overline{x}^0=-\overline{x}^-$, $L_-\overline{x}^-=0$
and of the Leibniz rule for $L_-$ we find that 
$$
L_-(\overline{x}^+)^l=\underbrace{\overline{x}^0\left(\overline{x}^+\right)^{l-1}
+\overline{x}^+\overline{x}^0\left(\overline{x}^+\right)^{l-2}
+\cdots+\left(\overline{x}^+\right)^{l-1}\overline{x}^0}_{l \mbox{ monomials in }\overline{x}^+,\overline{x}^0,\overline{x}^{-1}},
$$
and more generally that $L_-^h(\overline{x}^+)^l$ can be written as the sum $\sum'$ of  $n\le l^h$ 
(not necessarily distinct) monomials in $\overline{x}^+,\overline{x}^0,\overline{x}^{-1}$ with coefficients $\pm 1$. 
This homogenous polynomial, and therefore also $\widehat{Y}_l^m$, is completely symmetric with respect to
permutations of the factors $\overline{x}^a$, because both the monomial 
$(\overline{x}^+)^l$ and the Leibniz rule for $L_-$ are. 
The same occurs with $L_-^h(t^+)^l$. Hence
\be
Y_l^m-\widehat{Y}_l^m=R_l^m\hspace{0.1cm}\underbrace{{\sum}' \pm\left(
t^{a_1}t^{a_2}\cdots t^{a_l}-\overline{x}^{a_1}\overline{x}^{a_2}\cdots \overline{x}^{a_l}\right)}_{n\le
l^{l-m}\mbox{ terms }}, \qquad R_l^m:=M_l\sqrt{\frac{(l+m)!2^{l-m}}{(2l)!(l-m)!}}.             \label{differenceYlm}
\ee

As a fuzzy analog  of the vector space $B(S^2)$ we adopt
\be
{\cal C}_\Lambda:=\left\{\sum_{l=0}^{2\Lambda}\sum_{m=- l}^l
f_l^m \widehat{Y}_l^m\:,\: f_l^m\in\CC\right\}
\subset\A_\Lambda\subset B[{\cal L}^2(S^2)];
\label{def_CLambda3D}
\ee
here the highest $l$ is $2\Lambda$ because $(\overline{x}^{+})^{2\Lambda}\propto \widehat{Y}_{2\Lambda}^{2\Lambda}$ is the 
highest power of $\overline{x}^{+}$ acting nontrivially on $\Hi_\Lambda$ (it does 
not annihilate $\psi_\Lambda^{-\Lambda}$). By construction 
\be
\qquad\qquad {\cal C}_\Lambda=\bigoplus\limits_{l=0}^{2\Lambda} V_l,\qquad\qquad V_l:=
\left\{\sum_{m=- l}^l f^m \widehat{Y}_l^m\:,\: f^m\in\CC\right\} \label{deco2}
\ee
is the decomposition
of ${\cal C}_\Lambda$ into irreducible components under $O(3)$. $V_l$ is trace-free
for all $l>0$, i.e. its projection on the singlet component $V_0$ is zero. (\ref{deco2}) becomes the
decomposition of $B(S^2),C(S^2)$ in the limit $\Lambda\to\infty$. 
As a fuzzy analog  of $f\in B(S)$
we adopt
\be
\hat f_\Lambda:=\sum_{l=0}^{2\Lambda}\sum_{|m|\leq l}f_l^m 
\widehat{Y}_l^m\in\A_\Lambda\subset B[{\cal L}^2(S^2)].
\ee

In appendix \ref{convergD3} we
prove first that the operators $\overline{x}^a$ converge strongly to $t^{a}$  as $\Lambda\to\infty$
if we choose $k(\Lambda)$ fulfilling   (\ref{consistency3D}). Again, since
for all $\Lambda\!>\!0$ the operator $\overline{x}^a$ annihilates $\Hi_\Lambda^\perp$, 
$\overline{x}^a$  {\it does not} converge  to $t^a$ {\it in operator norm}.
Then  we prove the more general 

\begin{propo} \label{propoD3}
If we choose $k(\Lambda)\ge 2^{3\Lambda+3}\Lambda ^{\Lambda+5}(\Lambda\!+\!1)$, then for all $f,g\in B(S^2)$ the following strong limits as $\Lambda\rightarrow \infty$ hold: $\hat{f}_{\Lambda}\rightarrow f\cdot,\widehat{\left(fg\right)}_{\Lambda}\rightarrow fg$ and $\hat{f}_{\Lambda}\hat{g}_{\Lambda}\rightarrow fg\cdot$. 
\end{propo}
\noindent
The last statement says that the product in $\mathcal{A}_{\Lambda}$ of the approximations $\widehat{f}_{\Lambda}$,$\widehat{g}_{\Lambda}$ goes to the product in $B\left[\mathcal{L}^2(S)\right]$ of $f\cdot,g\cdot$.

The above dependence of $k$ on $\Lambda$ 
is by no means optimal, i.e. better estimates will presumably allow to prove the same result
with a function $k(\Lambda)$ growing much less rapidly.

\section{Final remarks, outlook and conclusions}
\label{Conclu}

For both dimensions \ $d=1,2$ \ we have introduced a finite-dimensional 
approximation of quantum mechanics on the sphere  $S^d_{\Lambda}$
 by projecting below a suitable energy cutoff $\overline{E}$ quantum mechanics of  a  particle in $\RR^{D}$ ($D=d\!+\!1$) configuration space
subject to a rotation invariant potential $V(r)$  with a very sharp minimum on the sphere of  radius \ $r=1$. \ By parametrizing both 
the confining parameter $k$ and $\overline{E}$ by a positive integer $\Lambda$ we have
obtained a sequence of $O(D)$-equivariant such approximations.
The following common features emerge.  The algebra of observables 
$\A_\Lambda$ on the  Hilbert space $\Hi_{\Lambda}$ 
is isomorphic to $\pi_\Lambda[Uso(D\!+\!1)]$, 
where  $\pi_\Lambda$ is a suitable irreducible unitary representation
of $U\!so(D\!+\!1)$ on $\Hi_{\Lambda}$. 
On the other hand,
$\Hi_{\Lambda}$ carries 
a {\it reducible} representation of the subalgebra 
\ $U\!so(D)$ \ generated by the projected angular momentum components
$\overline{L}_{ij}$, more precisely the direct sum
of {\it all} irreducible representations fulfilling the cutoff condition $L^2\le \Lambda(\Lambda+d-1)$; 
a similar
 decompostion holds for the subspace $\C_\Lambda\subset\A_\Lambda$ 
of completely symmetrized polynomials in the projected coordinates $\overline{x}^i$. 
In the $\Lambda\to\infty$  limit
these become the decompositions of  the Hilbert space ${\cal L}^2(S^d)$ and of the
algebra of operators $C(S^d)$ acting on ${\cal L}^2(S^d)$, respectively. 
The  $\overline{x}^i$, or
alternatively the elements $X^i$ of a corresponding basis of \ $so(D\!+\!1)\setminus so(D)$, \  generate the algebra of observables $\A_\Lambda\simeq \pi_\Lambda[Uso(D\!+\!1)]$
[the relation between them is of the form $\overline{x}^i=g(L^2)X^ig(L^2)$]; 
their commutators span 
$so(D)$, the Lie algebra of the angular momentum components $L_{ij}$, as in the Snyder algebra; 
a basis of $\A_\Lambda$ is made up of $\pi_\Lambda$-images of mononomials - with a fixed ordering -
in the elements of a basis of $so(D\!+\!1)$ , by the Poincar\'e-Birkhoff-Witt theorem.
The square distance $\R^2=\overline{x}^i\overline{x}^i$ 
from the origin is not identically 1, but a function of $L^2$ with spectrum close to 1. 
The whole construction is $SO(D\!+\!1)$- and $O(D)\subset  SO(D\!+\!1)$-equivariant.
We naturally embed $\Hi_\Lambda$ in ${\cal L}^2(S^d)$ and $\A_\Lambda$ in 
the algebra ${\cal O}(S^d)$ of operators on ${\cal L}^2(S^d)$.
In the $\Lambda\to\infty$ limit we recover ordinary $O(D)$-equivariant quantum mechanics on ${\cal L}^2(S^d)$,
with the observables   $\overline{L}_{ij}$, $\overline{x}^i$ going to the angular momentum components $L_{ij}$ and to the coordinates $x^i$ of $S^d$   configuration space, because $\R^2\to 1$ and
\ $\Hi_\Lambda,\A_\Lambda$  respectively ``invade'' ${\cal L}^2(S^d)$ and the whole ${\cal O}(S^d)$. In particular, every element $f$ of $B(S^d)$ or $C(S^d)$ is 
the strong limit of a sequence $\{\hat f_\Lambda\}$ of  elements $\hat f_\Lambda\in\C_\Lambda$.

Our approach seems easily applicable  with the same features to higher dimensions,
where comparison with previous proposals is possible. 
The fuzzy spheres of dimension $d\ge 3$ of \cite{GroKliPre96,Ramgoolam,Dolan:2003th} are 
based on the algebra $End(V)$ of endomorphisms of the carrier space $V$ of
a particular {\it irreducible} representation of $SO(d+1)$, so that 
the square distance from the origin $\R^2$ be  central and can be set strictly equal to 1. The commutation relations
are also of the Snyder type (although presumably slightly different from ours), hence equivariant with respect to the full group $O(d+1)$. 
In \cite{Ste16,Ste17} Steinacker and Sperling consider the possibility of a fuzzy 4-sphere $S^4_N$
with  a reducible representation of $Uso(5)$ on a Hilbert space $V$ obtained decomposing an irreducible representation $\pi$ of $Uso(6)$ characterized by a triple of highest weights
$(N,n_1,n_2)$; so $End(V)\simeq  \pi[Uso(6)]$, in analogy with our scheme. 
The elements $X^i$ of a basis of \ 
$so(6)\setminus so(5)$ \ play the role of noncommutative cartesian coordinates.
As a consequence the $O(5)$-scalar $\R^2=X^iX^i$ is no longer central, but its spectrum is still very close to 1  if $N\gg n_1,n_2$ [because 
then the decomposition of $V$ contains few irreducible representations under $SO(5)$]; 
note that in our approach this is guaranteed  by adopting
suitable $\overline{x}^i=g(L^2)X^ig(L^2)$ rather than $X^i$ as noncommutative cartesian coordinates. 
If $n_1=n_2=0$ then $\R^2\equiv 1$, and one recovers the  fuzzy 4-sphere \cite{GroKliPre96}. 
Their physical interpretation of  $End(V)$ is 
that it represents a fuzzy approximation of some fibre bundle on a sphere $S^4$ 
(rather than of the algebra  of observables of a quantum particle on a $S^4$).
If one wishes to describe a scalar field on such fuzzy   $S^4$, or more generally $S^d$ with $d\ge 3$, one can 
project out the unwanted modes of \cite{lam}, but this makes the product of spherical harmonics
non-associative. Alternatively, in \cite{Medina:2002pc,Dolan:2003th} unwanted modes are only suppressed in probability 
(not completely
eliminated) in the path-integral of the quantum field theory by adding suitable kinetic terms in the action.
Starting from fields on fuzzy $\CC P^1\simeq S^2$,  in \cite{Dolan:2003kq}  this idea is used also to  introduce an effective quantum field theory on a fuzzy $S^1$ 
by adding suitable kinetic terms in the action that suppress the modes $Y^l_m$ with $l<\Lambda$;
 the remaining $Y^\Lambda_m$, $|m|\le \Lambda$, play the role  of $u^m=e^{im\varphi}$ as elements of 
a basis of a fuzzy circle.

\subsubsection*{Acknowledgments}

We are grateful to F. D'Andrea and T. Weber for useful discussions.
We acknowledge partial support by  COST Action MP1405  {\it Quantum Structure of Spacetime}.

\section{Appendix}
\label{Appendix}

We shall repeatedly use the formula
\bea
\int_{-\infty}^{+\infty}e^{-a\rho^2+b\rho}\rho^n d\rho=e^{b^2/4a}\sqrt{\frac{\pi}{a}}\:\sum_{h=0}^{n/2}
\left(\!\!\begin{array}{c} n\\ 2h
\end{array}\!\!\right)
\left(\frac{b}{2a}\right)^{n-2h}\frac{(2h-1)!!}{(2a)^h}                \label{esp}
\eea
(for all $a>0$ and $b\in\mathbb{R}$), which can be easily derived from $\int_{-\infty}^{+\infty}e^{-z^2}dz=\sqrt{\pi}$
through integration by parts and a linear change of integration variable.

\subsection{Calculation of a rather general scalar product in $D=2$}

As a preliminary step,  we prove a formula regarding a matrix element of a general form.
\begin{propo} 
For every entire function $f(\rho)$ not depending on $k$ 
and  $h\!\in\!\mathbb{Z}$ the following asymptotic expansion in $1/\sqrt{k}$  holds
\bea
&&T_{m,m'}^f:=\langle\psi_{m'}, f(\rho)e^{ih\varphi}\psi_m\rangle=\delta^h_{m'-m}\,K_{m,m'}
\left[ \exp\left(\frac {\partial_\rho^2}{4c_{m,m'}}\right) f (\rho) \right]_{\rho=\rho_{m,m'}}\label{flemma}\\[10pt]
&&  \mbox{where }\ba{l}
\displaystyle c_{m,m'}:=\frac{\sqrt{k_m}\!+\!\sqrt{k_{m'}}}2=
\sqrt{2k}\left[1\!-\!\frac{1}{\sqrt{2k}}\!+\!\frac{3\!-\!m^2\!-\!m'{}^2}{4k} \right]+O\left(\frac 1{k}\right), \\[12pt]
 \displaystyle \rho_{m,m'}:=
\frac{2\!+\!\sqrt{k_m}\widetilde{\rho}_m\!+\!\sqrt{k_{m'}} \widetilde{\rho}_{m'}}{2c_{m,m'} }=
\frac{2}{\sqrt{2k}}\!+\!\frac{m^2\!+\!m'{}^2\!+\!2}{4k}+O\left(\frac 1{k^{3/2}}\right), \\[12pt]
\displaystyle K_{m,m'}:=\sqrt{\!\frac{4\pi^3}{c_{m,m'}}} N_m\overline{N_{m'}} 
e^{\frac{\left[2+\sqrt{k_m}\widetilde{\rho}_m+\sqrt{k_{m'}} \widetilde{\rho}_{m'}
\right]^2}{4c_{m,m'}}-\frac{\sqrt{k_m}\widetilde{\rho}_m^2+\sqrt{k_{m'}} \widetilde{\rho}_{m'}^2}2}=1+O\left(\frac{1}{k^{\frac{3}{2}}}\right),
\ea \label{deffs}
\eea
where $N_m=\sqrt[4]{\frac{\sqrt{k_m}}{4\pi^3}}e^{-\frac 1{2\sqrt{k_m}} - \widetilde{\rho}_m}$  is the normalization  of $\psi_m$
(up to a phase).
The powers in $1/\sqrt{k}$ arise from the Taylor expansion of the exponential and the definition of $c_{m,m'}$, e.g. 
$$
T_{m,m'}^f=f\!\left(\rho_{m,m'}\right) +\frac {f''\!\left(\rho_{m,m'}\right)}{4c_{m,m'}} 
+\frac {f^{(4)}\!\left(\rho_{m,m'}\right)}{32 (c_{m,m'})^2} +O\left(\frac 1 {k^{3/2}}\right).
$$
If $f(\rho)$ has continuous derivatives up to order $2h+1$ the above asymptotic expansion holds up to order $2h$.
\end{propo}

As consequences, setting $a:=1\!+\!\frac 94\frac{1}{\sqrt{2k}}\!+\!\frac{137}{64k}$, we find up to terms $O\left(1 /k^{3/2}\right)$
\bea
&& \sqrt{2}\langle\psi_{m'},u^{m'-m}\psi_m\rangle=\langle\psi_{m'}, e^{i(m'-m)\varphi}\psi_m\rangle=K_{m,m'}=1,\\
&& \langle\psi_{m'}, e^{n\rho+i(m'-m)\varphi}\psi_m\rangle=K_{m,m'}e^{n\rho_{m,m'}+\frac {n^2}{4c_{m,m'}}},
\\
&& \langle\psi_{m+1},x^+\psi_m\rangle=\frac{\langle\psi_{m+1}, e^{\rho+i\varphi}\psi_m\rangle}{\sqrt{2}}=
\frac{K_{m,m+1}}{\sqrt{2}}e^{\rho_{m,m+1}+\frac {1}{4c_{m,m+1}}}= \frac{1}{\sqrt{2}}\left[1\!+\!\frac{9}{4\sqrt{2k}}\right.\nn[4pt]
&&\quad\quad +\left.\frac{m(m\!+\!1)\!+\!137/32}{2k}\right]=
\frac{a}{\sqrt{2}}\left(1+\frac{m(m+1)}{2k}\right)=\frac{a}{\sqrt{2}}\sqrt{1+\frac{m(m+1)}{k}},\\
&& \langle\psi_{m-1},x^-\psi_m\rangle=
\overline{\langle\psi_m,x^+\psi_{m-1}\rangle}=\frac{1}{\sqrt{2}}\left[1\!+\!\frac{9}{4\sqrt{2k}}\!+\!\frac{m(m\!-\!1)\!+\!137/32}{2k}\right]  \nn
&&\quad\quad\qquad\qquad\quad =\frac{a}{\sqrt{2}}\sqrt{1+\frac{m(m-1)}{k}},\\
&&|\langle\psi_{m\pm 1},x^\pm\psi_m\rangle|^2=\frac{a^2}{2}\left(1+\frac{m(m\pm 1)}{k}\right).
\eea

\bigskip
{\bf Proof of the proposition.} \ From (\ref{psi_m}) we find
\bea
&& T_{m,m'}^f= \int_0^{2\pi}\!\!\! 
d\varphi \int_0^{+\infty}\!\!\!\! dr\, r \,\overline{\psi_{m'}(r,\varphi)} f(\rho)e^{ih\varphi}\psi_m(r,\varphi)\nn
&& =N_m\overline{N_{m'}}\, 2\pi\delta^h_{m'-m}\,
\int_{-\infty}^{+\infty}\!\!\!\!d\rho\,f(\rho) \,e^{2\rho-\left(\rho-\widetilde{\rho}_m\right)^2\frac{\sqrt{k_m}}2
-\left(\rho-\widetilde{\rho}_{m'}\right)^2\frac{\sqrt{k_{m'}}}2}\nn
&& =N_m\overline{N_{m'}}\, 2\pi\delta^h_{m'-m}\, \int_{-\infty}^{+\infty}\!\!\!\!d\rho\,f(\rho) \,
e^{ -\frac{\sqrt{k_m}+\sqrt{k_{m'}}}2\rho^2
+\rho \left[2+\sqrt{k_m}\widetilde{\rho}_m+\sqrt{k_{m'}} \widetilde{\rho}_{m'}
\right] -\frac{\sqrt{k_m}\widetilde{\rho}_m^2+\sqrt{k_{m'}} \widetilde{\rho}_{m'}^2}2 }\nn
&& =N_m\overline{N_{m'}}\, 2\pi\delta^h_{m'-m}\, e^{\frac{\left[2+\sqrt{k_m}\widetilde{\rho}_m+\sqrt{k_{m'}} \widetilde{\rho}_{m'}
\right]^2}{2(\sqrt{k_m}+\sqrt{k_{m'}})}-\frac{\sqrt{k_m}\widetilde{\rho}_m^2+\sqrt{k_{m'}} \widetilde{\rho}_{m'}^2}2} \int_{-\infty}^{+\infty}\!\!\!\!d\rho\,f(\rho) \,
e^{ -\frac{\sqrt{k_m}+\sqrt{k_{m'}}}2(\rho-\rho_{m,m'})^2 }\nn
&& =N_m\overline{N_{m'}}\, 2\pi\delta^h_{m'-m}\, e^{\frac{\left[2+\sqrt{k_m}\widetilde{\rho}_m+\sqrt{k_{m'}} \widetilde{\rho}_{m'}
\right]^2}{4c_{m,m'}}-\frac{\sqrt{k_m}\widetilde{\rho}_m^2+\sqrt{k_{m'}} \widetilde{\rho}_{m'}^2}2}
 \int_{-\infty}^{+\infty}\!\!\!\!dz\,e^{-z^2c_{m,m'}} \,f\!\left(\!\rho_{m,m'}\!+\!z\!\right);\label{inter}
\eea
in the last step we have changed the integration variable \ $\rho\!\mapsto\! z=\rho\!-\!\rho_{m,m'}$.
Using the Taylor expansion
of $f\!\left(\!\rho_{m,m'}\!+\!z\!\right) $ and the vanishing of integrals of odd functions over $\mathbb{R}$ we find
\bea
 && \int\limits_{-\infty}^{+\infty}\!\!\! dz\,e^{-z^2c_{m,m'}}  f\!\left(\!\rho_{m,m'}\!+\!z\!\right) =\!\!\!
 \int\limits_{-\infty}^{+\infty}\!\!\! dz\,e^{-z^2c_{m,m'}} \sum_{n=0}^\infty  f^{(n)}\!\left( \rho_{m,m'}\right) \frac {z^n}{n!}\nn
&& = \sum_{n=0}^\infty f^{(2n)}\!\left( \rho_{m,m'}\!\right)\!\!\int\limits_{-\infty}^{+\infty}\!\!\! dz\,e^{-z^2c_{m,m'}}  \frac {z^{2n}}{(2n)!}= \sum_{n=0}^\infty \frac{f^{(2n)}\!\left( \rho_{m,m'}\!\right)}{(c_{m,m'})^{n+1/2}}
\int\limits_{-\infty}^{+\infty}\!\!\! dy\,e^{-y^2}  \frac {y^{2n}}{(2n)!}\nn
&& = \sqrt{\frac{\pi}{c_{m,m'}}}\!\sum_{n=0}^\infty \!\frac{f^{(2n)}\!\left( \rho_{m,m'}\right) }{(4c_{m,m'})^n \, n!}
= \sqrt{\frac{\pi}{c_{m,m'}}}\sum_{n=0}^\infty\! \left. \frac{\partial_\rho^{2n} }{(4c_{m,m'})^n \, n!}\, f(\rho)\right\vert_{\rho= \rho_{m,m'}}\nn
&& = \sqrt{\frac{\pi}{c_{m,m'}}}\left.\exp\left[ \frac{\partial_\rho^2 }{4c_{m,m'}}\right] f(\rho)\right\vert_{\rho= \rho_{m,m'}}
;\nonumber
\eea
here we have used the identity $\int_{-\infty}^{+\infty}\!\!\! dy\,e^{-y^2}y^{2n}=\sqrt{\pi} (2n\!-\!1)!!/2^n$,
which can be proved iterating integration by parts. Replacing in (\ref{inter}) we find  (\ref{flemma}), with
\be
 K_{m,m'}:=\sqrt{\!\frac{4\pi^3}{c_{m,m'}}} N_m\overline{N_{m'}} e^{\frac{\left[2+\sqrt{k_m}\widetilde{\rho}_m+\sqrt{k_{m'}} \widetilde{\rho}_{m'}
\right]^2}{4c_m^h}-\frac{\sqrt{k_m}\widetilde{\rho}_m^2+\sqrt{k_{m'}} \widetilde{\rho}_{m'}^2}2} \label{defTK}
\ee
In particular choosing $h=0$, $f\equiv 1$ and recalling the normalization condition $T_{m,0}^1=\Vert\psi_m\Vert^2=1$
we determine the normalization factors $N_m$:
$$
|N_m|^2\sqrt{\frac{4\pi^3}{\sqrt{k_m}}}e^{\frac 1{\sqrt k_m} +2\widetilde{\rho}_m}=1\qquad \Longrightarrow \qquad
N_m=\sqrt[4]{\frac{\sqrt{k_m}}{4\pi^3}}e^{-\frac 1{2\sqrt{k_m}} - \widetilde{\rho}_m}
$$
Hence,
\bea
&&\sqrt{\!\frac{4\pi^3}{c_{m,m'}}}N_m\overline{N_{m'}}
=\frac {\sqrt{2}\sqrt{\sqrt[4]{k_mk_{m'}}} }{\sqrt{\sqrt{k_m}+\sqrt{k_{m'}}}} 
e^{-\frac 1{2\sqrt {k_m}} - \widetilde{\rho}_m -\frac 1{2\sqrt{k_{m'}}} - \widetilde{\rho}_{m'}}\nn
&&=\frac {\sqrt{2}}{\sqrt{\sqrt[4]{\frac{k_m}{k_{m'}}} +\sqrt[4]{\frac{k_{m'}}{k_m}}} }
e^{-\frac 1{2\sqrt {k_m}} - \widetilde{\rho}_m -\frac 1{2\sqrt{k_{m'}}} - \widetilde{\rho}_{m'}}=
e^{-\frac 1{2\sqrt {k_m}} - \widetilde{\rho}_m -\frac 1{2\sqrt{k_{m'}}} - \widetilde{\rho}_{m'}}
\nonumber
\eea
which replaced in (\ref{defTK}) gives (\ref{deffs})$_3$.  We now determine $c_{m,m'},\rho_{m,m'},K_{m,m'}$ at lowest order.
By (\ref{krho_m}), up to $O(1/k^{\frac 32})$,
\bea
&& \sqrt{k_m}=\sqrt{2k}\left[1\!-\!\frac{1}{\sqrt{2k}}\!+\! \frac{\frac{3}{2}\!-\!m^2}{2k}\right]\!,\qquad \sqrt{k_m}\widetilde{\rho}_m=1+\frac{m^2-1}{\sqrt{2k}}+\frac{3-4m^2}{4k}\nn 
&&\frac{1}{\sqrt{k_m}}=\frac{1}{\sqrt{2k}}+\frac{1}{2k}\!,\!\frac{1}{\sqrt{k_m}+\sqrt{k_{m'}}}=\frac{1}{\sqrt{2k}}\left(\frac{1}{2}+\frac{1}{2\sqrt{2k}} \right)\!,\nn
&& c_{m,m'}=\frac{\sqrt{k_m}\!+\!\sqrt{k_{m'}}}2
=\sqrt{2k}\left[1\!-\!\frac{1}{\sqrt{2k}}\!+\!\frac{3\!-\!m'{}^2\!-\!m^2}{4k} \right],\nn
&&\rho_{m,m'}=\frac{2+\sqrt{k_m}\widetilde{\rho}_m+\sqrt{k_{m'}}\widetilde{\rho}_{m'}}{2c_{m,m'}}=\frac{2}{\sqrt{2k}}+\frac{m^2+m'^2+2}{4k}.\nn
\nonumber
\eea
then using (\ref{defs}) and by algebraic manipulations one has
\bea
&K_{m,m'}&=\sqrt{\frac{2}{\sqrt[4]{\frac{k_m}{k_{m'}}}+\sqrt[4]{\frac{k_{m'}}{k_m}}}}\cdot e^{\frac{\left[2+\sqrt{k_m}\widetilde{\rho}_m+\sqrt{k_{m'}}\widetilde{\rho}_{m'} \right]^2}{2\left(\sqrt{k_m}+\sqrt{k_{m'}}\right)}-\frac{\sqrt{k_m}\widetilde{\rho}_m^2+\sqrt{k_{m'}}\widetilde{\rho}^2_{m'}}{2}-\frac{1}{2\sqrt{k_m}}-\frac{1}{2\sqrt{k_{m'}}}-\widetilde{\rho}_m-\widetilde{\rho}_{m'}}\nn
&&=\sqrt{\frac{2}{\sqrt[4]{\frac{k_m}{k_{m'}}}+\sqrt[4]{\frac{k_{m'}}{k_m}}}}\cdot e^{-\frac{\sqrt{k_mk_{m'}}}{2\left(\sqrt{k_m}+\sqrt{k_{m'}} \right)}\left(\frac{E'_m+\sqrt{k_m}}{k_m}-\frac{E'_{m'}+\sqrt{k_{m'}}}{k_{m'}} \right)^2}\nn.
&&=\sqrt{\frac{2}{\sqrt[4]{\frac{k_m}{k_{m'}}}+\sqrt[4]{\frac{k_{m'}}{k_m}}}}\cdot e^{-\frac{\sqrt{k_mk_{m'}}}{2\left(\sqrt{k_m}+\sqrt{k_{m'}} \right)}\left(\widetilde{\rho}_m+\frac{1}{\sqrt{k_m}}-\widetilde{\rho}_{m'}-\frac{1}{\sqrt{k_{m'}}} \right)^2}\nn.
\nonumber
\eea
and since
\bea
&&\frac{\sqrt{k_mk_{m'}}}{2\left(\sqrt{k_m}+\sqrt{k_{m'}} \right)}\left(\widetilde{\rho}_m+\frac{1}{\sqrt{k_m}}-\widetilde{\rho}_{m'}-\frac{1}{\sqrt{k_{m'}}} \right)^2=0,\nn
&& \frac{k_m}{k_{m'}}=1+\frac{m'{}^2\!-\!m^2}{k},\qquad \sqrt[4]{\frac{k_m}{k_{m'}}}=1+\frac {m'{}^2\!-\!m^2}{4k},
\qquad\sqrt[4]{\frac{k_m}{k_{m'}}} + \sqrt[4]{\frac{k_{m'}}{k_m}}=2,\qquad \label{inter1}\nonumber
\eea
one has $K_{m,m'}=1$, up to $O\left(k^{-\frac{3}{2}}\right)$. \hfill
\qedsymbol

\subsection{Calculation of the action of operators $\overline{\partial_{\pm}}$ in $D=2$}

We first compute $\overline{\partial}_+\psi_m$:
\bea
\hspace{-1.3cm}\overline{\partial}_+\psi_m=\sum_h\psi_h \langle\psi_h,\overline{\partial}_+\psi_m\rangle=
\left\{\ba{ll}
 \psi_{m-1}\, \langle\psi_{m-1}, \partial_+\psi_m\rangle\quad &
\mbox{ if }-\Lambda+1\leq m\leq \Lambda \\[8pt]
0 & \mbox{otherwise.}
\ea\right.
\label{barde+}
\eea
By (\ref{mpartial}) we can calculate this scalar product using (\ref{flemma}) with $h=-1$ and 
$$
f\left(\rho\right)=\frac{e^{-\rho}}{2}\left[m-\left(\rho-\widetilde{\rho}_m\right)\sqrt{k_m}\right];
$$
consequently,
$$
\langle\psi_{m-1},\partial_+\psi_m\rangle=\frac{1}{\sqrt{2}}\left(m-\frac{1}{2}-\frac{\frac{3}{4}m-\frac{3}{8}}{\sqrt{2k}}-\frac{m^3-\frac{3}{2}m^2+\frac{79}{32}m-\frac{63}{64}}{2k}\right)+O\left(\frac{1}{k^{\frac{3}{2}}}\right).
$$
From (\ref{barde+}) we have
\be
\overline{\partial}_+\psi_m=
\left\{
\ba{ll}
\frac{1}{\sqrt{2}}\left(b+m-\frac{3m}{4\sqrt{2k}}-\frac{m^3-\frac{3}{2}m^2+\frac{79}{32}m}{2k}\right)\psi_{m-1}&\mbox{if}-\Lambda+1\leq m\leq \Lambda\\
0&\mbox{otherwise.}\\
\ea
\right.
\label{part+}
\ee
where $b=-\frac{1}{2}+\frac{3}{8\sqrt{2k}}+\frac{63}{128k}$. Moreover
$$
\overline{\partial}_-\psi_{m-1}=\sum_{h}{\psi_h\langle\psi_h,\partial_-\psi_{m-1}\rangle}=
\left\{
\ba{ll}
\psi_m\langle\psi_m,\partial_-\psi_{m-1}\rangle&\mbox{if}-\Lambda+1\leq m\leq \Lambda\\
0&\mbox{otherwise.}\\
\ea
\right.$$
\bea
\langle\psi_m,\partial_-\psi_{m-1}\rangle&=&\langle\partial_+\psi_m,\psi_{m-1}\rangle =
\overline{\langle\psi_{m-1},\partial_+\psi_m\rangle}\nn
&=&\frac{1}{\sqrt{2}}\left(b+m-\frac{3m}{4\sqrt{2k}}-\frac{m^3-\frac{3}{2}m^2+\frac{79}{32}m}{2k}\right)+O\left(\frac{1}{k^{\frac{3}{2}}}\right) \nonumber
\eea
whence (replacing $m$ with $m+1$)
\be
\overline{\partial}_-\psi_m=
\left\{\!\!
\ba{ll}
\frac{1}{\sqrt{2}}\left(b\!+\!m\!+\!1\!-\!\frac{3(m\!+\!1)}{4\sqrt{2k}}\!-\!\frac{(m\!+\!1)^3\!-\!\frac{3}{2}(m\!+\!1)^2\!+\!\frac{79}{32}(m\!+\!1)}{2k}\right)\psi_{m+1}&\mbox{if}-\Lambda\leq m\leq \Lambda-1,\\
0&\mbox{otherwise.}
\ea
\right.
\label{part-}
\ee
Eq. (\ref{part+}), (\ref{part-}) imply at leading order
\bea
&&\overline{\partial}_+\overline{\partial}_-\psi_m=
\left\{\!\!
\ba{ll}
\frac{1}{2}\left(b+m+1-\frac{3(m+1)}{4\sqrt{2k}}-\frac{(m+1)^3-\frac{3}{2}(m+1)^2+\frac{79}{32}(m+1)}{2k}\right)^2\psi_{m}&\mbox{if}-\Lambda\leq m\leq \Lambda-1,\\
0&\mbox{otherwise,}
\ea
\right. \nn
&&\overline{\partial}_-\overline{\partial}_+\psi_m=
\left\{\!\!
\ba{ll}
\frac{1}{2}\left(b+m-\frac{3m}{4\sqrt{2k}}-\frac{m^3-\frac{3}{2}m^2+
\frac{79}{32}m}{2k}\right)^2\psi_m&\mbox{if}-\Lambda+1\leq m\leq \Lambda,\\
0&\mbox{otherwise,}
\ea
\right. \nonumber
\eea
whence
\be
\left[\overline{\partial}_+,\overline{\partial}_-\right]\psi_m=\left\{\! 
\ba{ll}
\left(m\!-\!\frac{3m}{2\sqrt{2k}}\!-\!\frac{4m^3\!+\!\frac{31}{8}m}{2k}\right)\psi_m&\mbox{if }|m|\leq \Lambda-1,\\
-\frac{1}{2}\left(b\!+\!m\!-\!\frac{3m}{4\sqrt{2k}}\!-\!\frac{m^3\!-\!\frac{3}{2}m^2\!+\!\frac{79}{32}m}{2k}\right)^2\psi_{m}&\mbox{if }m=\Lambda,\\
\frac{1}{2}\left(b\!+\!m\!+\!1\!-\!\frac{3(m\!+\!1)}{4\sqrt{2k}}\!-\!\frac{(m\!+\!1)^3\!-\!\frac{3}{2}(m\!+\!1)^2\!+\!\frac{79}{32}(m\!+\!1)}{2k}\right)^2\psi_{m}&\mbox{if }m=-\Lambda,\\
0&\mbox{otherwise.}
\ea
\right.
\ee
and the analogous of laplacian
\bea
\left(\overline{\partial}_+\overline{\partial}_-+\overline{\partial}_-\overline{\partial}_+\right)\psi_m=\left\{\!
\ba{ll}
\left(m^2\!+\!\frac{1}{4}\!-\!\frac{\frac{3m^2}{2}\!+\!\frac{3}{8}}{\sqrt{2k}}\!-\!\frac{2m^4\!+\!\frac{47m^2}{8}\!+\!\frac{27}{32}}{2k}
\right)\psi_m&\mbox{if }|m|\leq \Lambda-1,\\
\frac{1}{2}\left(b\!+\!m\!-\!\frac{3m}{4\sqrt{2k}}\!-\!\frac{m^3\!-\!\frac{3}{2}m^2\!+\!\frac{95}{32}m}{2k}\right)^2\psi_{m}&\mbox{if }m=\Lambda,\\
\frac{1}{2}\left(b\!+\!m\!+\!1\!-\!\frac{3(m\!+\!1)}{4\sqrt{2k}}\!-\!\frac{m^3\!+\!\frac{3}{2}m^2\!+\!\frac{95}{32}m\!+\!\frac{79}{32}}{2k}\right)^2\psi_{m}&\mbox{if }m=-\Lambda,\\
0&\mbox{otherwise.}
\ea
\right.
\eea

Then one can conclude that there exists $4$ polinomyals $P_1,P_2,Q_1,Q_2$ such that
\be
\left[\overline{\partial}_+,\overline{\partial}_-\right]=P_1(\overline{L})+Q_1(\overline{L})\left(\widetilde{P}_{\Lambda}-\widetilde{P}_{-\Lambda}\right)\hspace{0.5cm}\mbox{ and }\hspace{0.5cm}\overline{\partial}_+\overline{\partial}_-+\overline{\partial}_-\overline{\partial}_+=P_2(\overline{L})+Q_2(\overline{L})\left(\widetilde{P}_{\Lambda}-\widetilde{P}_{-\Lambda}\right).\label{derfine}
\ee

\subsection{Proof of proposition \ref{2Dconverge}}
\label{2Dconvergence}
Since
\bea
\eta^hu^m=u^{m+h}\alpha_{m,m+h},\qquad\mbox{where }\: \alpha_{m,n}:=\left\{\ba{ll}
\prod_{j=m}^{n-1}\sqrt{\!1\!+\!\frac{j(j\!+\!1)}{k}} & \mbox{ if }\: \Lambda\ge n>m\ge -\Lambda,  \\[8pt]
 1 \quad & \mbox{ if }\: \Lambda\ge n=m\ge -\Lambda,  \\[8pt]
\prod_{j=n}^{m-1}\sqrt{\!1\!+\!\frac{j(j\!+\!1)}{k}} & \mbox{ if }\:-\Lambda\le n<m\le \Lambda,  \\[8pt]
0& \mbox{ otherwise,}
\ea\right.
\eea
then, more explicitly,
\bea
\hat f_\Lambda \phi=\sum_{h=-2\Lambda}^{2\Lambda}f_h \eta^h\sum_{m=-\Lambda}^{\Lambda}\phi_mu^m
=\sum_{n=-\Lambda}^{\Lambda}u^n(\hat f_\Lambda \phi)_n,\qquad \mbox{ where }\: 
(\hat f_\Lambda \phi)_n:=\sum_{m=-\Lambda}^{\Lambda}f_{n-m}\phi_m\, \alpha_{m,n},
\nonumber
\eea
\bea
(f -\hat f_\Lambda)\phi=\sum_{n=-\Lambda}^{\Lambda}u^n \chi_n
+\sum_{|n|>\Lambda}u^n (f\phi)_n , \qquad \chi_n:=(f\phi)_n-(\hat f_\Lambda \phi)_n
\eea
[here $(f\phi)_n$ is the $n$-th Fourier coefficient of $f\phi\in{\cal L}^2(S)$], implying
\bea
 \Vert(f -\hat f_\Lambda)\phi\Vert^2=\sum_{n=-\Lambda}^{\Lambda}|\chi_n|^2
+\sum_{|n|>\Lambda}|(f\phi)_n|^2.        \label{norm1}
\eea
The second sum vanishes as $\Lambda\to\infty$. To show that the first sum
does as well we decompose
\bea
&& \chi_n=\sigma_n-\tau_n,\qquad
\sigma_n:=(f\phi)_n\!-\!\sum_{m=-\Lambda}^{\Lambda}f_{n-m}\phi_m,\quad
\tau_n:=(\hat f_\Lambda \phi)_n\!-\!\sum_{m=-\Lambda}^{\Lambda}f_{n-m}\phi_m\nn
&&\Rightarrow \qquad\qquad\sum_{n=-\Lambda}^{\Lambda} |\chi_n|^2\le 2\sum_{n=-\Lambda}^{\Lambda}\left(\left| \sigma_n\right|^2+\left| \tau_n\right|^2\right)
\eea
But
\bea
\sigma_n=\int_0^{2\pi}\frac{d\varphi}{2\pi}e^{-in\varphi}f(\varphi)\phi(\varphi)
\!-\!\sum_{m=-\Lambda}^{\Lambda}\phi_m\int_0^{2\pi}\frac{d\varphi}{2\pi}e^{-i(n-m)\varphi}f(\varphi)\nn
=\int_0^{2\pi}\frac{d\varphi}{2\pi}e^{-in\varphi}f(\varphi)\phi(\varphi)
\!-\!\int_0^{2\pi}\frac{d\varphi}{2\pi}e^{-in\varphi}f(\varphi)\sum_{m=-\Lambda}^{\Lambda}\phi_me^{im\varphi}\nn
=\int_0^{2\pi}\frac{d\varphi}{2\pi}e^{-in\varphi}f(\varphi)\left[\phi(\varphi)
\!-\! \phi_\Lambda (\varphi)\right]=\left( f\!\left[\phi\!-\! \phi_\Lambda \right]\right)_n \quad\Rightarrow\nn
\sum_{n=-\Lambda}^{\Lambda}\left| \sigma_n\right|^2 
\le \sum_{n\in\ZZ}\left|\left( f\!\left[\phi\!-\! \phi_\Lambda \right]\right)_n\right|^2=\left\Vert f\!\left[\phi\!-\! \phi_\Lambda \right]\right\Vert^2\le 
\left\Vert f\right\Vert_\infty^2\,
\left\Vert \phi -\phi_\Lambda \right\Vert^2\stackrel{\Lambda\to\infty}{\longrightarrow}0 \label{sigmalimit}
\eea
(we have used  $\left\Vert \phi -\phi_\Lambda \right\Vert \to 0$ as $\Lambda\to\infty$), and on the other hand
\bea
|\tau_n| & \le & \sum_{m=-\Lambda}^{\Lambda}|f_{n-m}|\, |\phi_m|\, \left[\alpha_{m,n}\!-\!1\right]
\le \sum_{m=-\Lambda}^{\Lambda} F \Phi \left[\left(\!1\!+\!\frac{\Lambda(\Lambda\!+\!1)}{k}\right)^{\frac{|n-m|}2}-1\right]\nn
& \le & F \Phi \sum_{m=-\Lambda}^{\Lambda}\left[\left(\!1\!+\!\frac{\Lambda(\Lambda\!+\!1)}{k}\right)^{\Lambda}-1\right]=F \Phi (2\Lambda\!+\!1)\left[\left(\!1\!+\!\frac{\Lambda(\Lambda\!+\!1)}{k}\right)^{\Lambda}-1\right]\nn
& \le &  F \Phi (2\Lambda\!+\!1)\left[e^{\frac{\Lambda^2(\Lambda\!+\!1)}{k}}-1\right] \nonumber
\eea
where $F=\max\limits_{m\in\ZZ}|f_m|$, $\Phi=\max\limits_{m\in\ZZ}|\phi_m|$, and we have used the inequality
$(1+y)^\Lambda< e^{y\Lambda}$ for all $y,\Lambda>0$.
Provided we choose $k(\Lambda)$ sufficiently large, e.g. $k\ge 2 {{\Lambda}}({{\Lambda}}\!+\!1)(2{{\Lambda}}\!+\!1)^2$, and note that \ $e^y\!-\!1<2y$  if  $0<y<1/2$, \ we thus find
$|\tau_n|\le F\Phi/(2{{\Lambda}}\!+\!1)$ and
\be
\sum_{n=-\Lambda}^{\Lambda}|\tau_n|^2\le \frac {F^2\Phi^2}{2\Lambda\!+\!1}\stackrel{\Lambda\to\infty}{\longrightarrow}0.                        \label{taulimit}
\ee
By (\ref{norm1}), (\ref{sigmalimit}), (\ref{taulimit})  we find 
\bea
\Vert(f -\hat f_\Lambda)\phi\Vert^2\le
\sum_{|n|>\Lambda}|(f\phi)_n|^2+2 \left\Vert f\right\Vert_\infty^2\,
\left\Vert \phi -\phi_\Lambda \right\Vert^2+ \frac {2F^2\Phi^2}{2\Lambda\!+\!1} \:
\stackrel{\Lambda\to\infty}{\longrightarrow}0,       \label{flimit}
\eea
i.e. $\widehat{f}_\Lambda\to f\cdot$ strongly for all $f\in B(S)$, as claimed. Replacing $f\mapsto fg$,
 we find also that $\widehat{(fg)}_\Lambda\to (fg)\cdot$ (strongly) for all $f,g\in B(S)$.
On the other hand, relation (\ref{flimit})  implies also 
\bea
\Vert(f -\hat f_\Lambda)\phi\Vert^2\le
\Vert f\phi\Vert^2+2 \left\Vert f\right\Vert_\infty^2\,
\left\Vert \phi -\phi_\Lambda \right\Vert^2+ \frac {2 \left\Vert f\right\Vert_\infty^2\Vert\phi\Vert^2}{2\Lambda\!+\!1} 
< 4 \left\Vert f\right\Vert_\infty^2\Vert\phi\Vert^2, \nn[6pt]
\Vert \hat f_\Lambda\phi\Vert 
\le \Vert (\hat f_\Lambda\!-\! f)\phi\Vert +\Vert f\phi\Vert 
\le \Vert (\hat f_\Lambda\!-\! f)\phi\Vert +\Vert f\Vert_\infty \Vert\phi\Vert\le 3 \Vert f\Vert_\infty \Vert\phi\Vert \nonumber
\eea
i.e. the operator norms $\Vert \hat f_\Lambda\Vert_{op}$ of the $ \hat f_\Lambda$ are uniformly bounded:
\be
\Vert \hat f_\Lambda\Vert_{op}\le 3 \Vert f\Vert_\infty.
\ee 
Therefore  (\ref{flimit}) implies also, as claimed,
\bea
\Vert(fg -\hat f_\Lambda\hat g_\Lambda)\phi\Vert
&\le & \Vert (f -\hat f_\Lambda)g\phi\Vert+\Vert \hat f_\Lambda(g-\hat g_\Lambda)\phi\Vert \nn[6pt]
&\le & \Vert (f -\hat f_\Lambda)(g\phi)\Vert+\Vert \hat f_\Lambda\Vert_{op} \: \: \Vert(g -\hat g_\Lambda)\phi\Vert
\stackrel{\Lambda\to\infty}{\longrightarrow}0.      \label{flimit'}
\eea

\subsection{Spherical Harmonics}
Let
$$
\left\{
\begin{array}{l}
x=r\sin{\theta}\cos{\varphi}\\
y=r\sin{\theta}\sin{\varphi}\\
z=r\cos{\theta}\\
\end{array}
\right.
\hspace{1cm}\mbox{ , }
\hspace{1cm}
\left\{
\begin{array}{l}
x^+=\frac{x+iy}{\sqrt{2}}=\frac{r\sin{\theta}e^{i\varphi}}{\sqrt{2}}\\
x^-=\frac{x-iy}{\sqrt{2}}=\frac{r\sin{\theta}e^{-i\varphi}}{\sqrt{2}}\\
x^0=z=r\cos{\theta}
\end{array}
\right.
,$$
$t^0=\frac{z}{r}$, $t^{+1}=\frac{x+iy}{\sqrt{2}r}$ and $t^-=\frac{x-iy}{\sqrt{2}r}$, then one has the following recurrence relations:
\bea
&&\hspace{-0.8cm}t^0 Y^m_l=\cos{\theta}Y^m_l=\sqrt{\frac{(l+m)(l-m)}{(2l+1)(2l-1)}}Y^m_{l-1}+\sqrt{\frac{(l+m+1)(l-m+1)}{(2l+1)(2l+3)}}Y^m_{l+1},\nn
&&\hspace{-0.8cm}t^{1} Y^m_l=\frac{\sin{\theta}e^{i\varphi}}{\sqrt{2}}Y^m_l=\frac{1}{\sqrt{2}}\left(\sqrt{\frac{(l\!-\!m)(l\!-\!m\!-\!1)}{(2l\!+\!1)(2l\!-\!1)}}Y^{m+1}_{l-1}-\sqrt{\frac{(l\!+\!m\!+\!1)(l\!+\!m\!+\!2)}{(2l\!+\!1)(2l\!+\!3)}}Y^{m+1}_{l+1}\right), \label{ClebschAB}\\
&&\hspace{-0.8cm}t^{-1} Y^m_l=\frac{\sin{\theta}e^{-i\varphi}}{\sqrt{2}}Y^m_l=\frac{1}{\sqrt{2}}\left(-\sqrt{\frac{(l\!+\!m)(l\!+\!m\!-\!1)}{(2l\!+\!1)(2l\!-\!1)}}Y^{m-1}_{l-1}+\sqrt{\frac{(l\!-\!m\!+\!1)(l\!-\!m\!+\!2)}{(2l\!+\!1)(2l\!+\!3)}}Y^{m-1}_{l+1}\right).\nonumber
\eea
The coefficients $A^{a,m}_l,B^{a,m}_l$ are related by 
\bea
&&A_l^{a,m}=\langle Y_{l-1}^{m+a},t^aY_l^m\rangle=\langle t^{-a} Y_{l-1}^{m+a},Y_l^m\rangle=\overline{\langle Y_l^m,t^{-a} Y_{l-1}^{m+a}\rangle};
=B_{l-1}^{-a,m+a}\label{AB}
\eea
and fulfill the properties (for all $l\ge 0$, $|m|\le l$)
\bea
&&A^{b,m}_lA^{-a,m+b+a}_{l}+A^{-b,m+b}_{l+1}A^{a,m+b}_{l+1}=A^{a,m}_lA^{-b,m+a+b}_{l}+A^{-a,m+a}_{l+1}A^{b,m+a}_{l+1},\nn[6pt]
&&A^{b,m}_{l+1}A^{a,m+b}_l=A^{a,m}_{l+1}A^{b,m+a}_l,
\qquad \qquad \sum_a A_{l+1}^{a,m}A_l^{-a,m+a}=0,
\label{Aproperties}\\
&&\sum_a(A_l^{a,m})^2=\frac l{2l\!+\!1},\qquad \sum_a(A_{l+1}^{a,m-a})^2=\frac {l\!+\!1}{2l\!+\!1},\qquad
\sum_a(A_l^{a,m})^2\!+\!\sum_a(A_{l+1}^{a,m-a})^2=1.\nonumber
\eea
Actually, the latter are equivalent to the identities \ $[t^a,t^b]=0$, \ $t^at^{-a}=1$ \  applied to \ $Y^m_l$. \ 

\subsection{Calculation of $\left\vert N_l\right\vert$ in  $D=3$}

$\left\vert N_l\right\vert$ can be determined setting $\langle \psi_{l}^m,\psi_{l}^m\rangle=1$ and we will choose 
$N_l=\left\vert N_l\right\vert$; using the orthonormality of the spherical harmonics, one obtains
\bea
&&1=|N_l|^2\int_0^{+\infty}\!\! e^{-(r-\widetilde{r}_l)^2\sqrt{k_l}}dr\simeq |N_l|^2\int_{-\infty}^{+\infty}\!\! e^{-(r-\widetilde{r}_l)^2\sqrt{k_l}}dr=|N_l|^2\sqrt{\frac{\pi}{\sqrt{k_l}}}\quad \Rightarrow \quad |N_l|=\frac{\sqrt[8]{k_l}}{\sqrt[4]{\pi}}\nn\label{appr1}
\eea
and the meaning of the approximation symbol $\simeq$ is explained in subsection \ref{Nlproof}.

\subsection{Calculation of a rather general scalar product in $D=3$}
\label{scp3D}

We compute the useful scalar product
\bea
\langle \psi_{L}^{m'},g(r)t^a\psi_l^m\rangle=\langle Y_{L}^{m'},t^aY_l^m\rangle\, 
\int_0^{+\infty}{f_l(r)f_L(r)g(r)dr}                               \label{genscp3}
\eea
with a generic $g(r)$ not depending on $k$; here we have used the decomposition 
$\psi_l^m\left(r,\theta,\varphi\right)=Y_l^m\left(\theta,\varphi\right)\frac{f_l(r)}{r}$ and factorized the integral over $\RR^3$ into
an integral over the angle variables and the integral on the radial one. 
By (\ref{tY'}) one finds
$$
\langle Y_{L}^{m'},t^aY_l^m\rangle\neq0\qquad \Leftrightarrow \qquad L=l\pm1 \mbox{ and }m'=m+a\nn
$$
The asymptotic expansion of the radial integral can be obtained from the general
formula
\bea
\int_0^{+\infty}{f_l(r)f_L(r)g(r)dr}= e^{-\frac{\sqrt{k_lk_L}\left(\widetilde{r}_l-\widetilde{r}_L\right)^2}{2\left(\sqrt{k_l}+\sqrt{k_L}\right)}}\:\sum_{n=0}^{+\infty}{ \frac{g^{(2n)}\left(\widehat{r}_{l,L}\right)}{(2n)!!\left(\sqrt{k_l}+\sqrt{k_L}\right)^n}},                          \label{general}
\eea
which we now prove:
\begin{align*}
\int_0^{+\infty}{f_l(r)f_L(r)g(r)dr}&=N_lN_L\int_0^{+\infty}{e^{-r^2\left(\frac{\sqrt{k_l}+\sqrt{k_L}}{2} \right)+r\left(\sqrt{k_l}\widetilde{r}_l+\sqrt{k_L}\widetilde{r}_L\right)-\frac{\sqrt{k_l}\widetilde{r}_l^2+\sqrt{k_L}\widetilde{r}_L^2}{2}}g(r)dr}\nn
&=N_lN_Le^{\frac{\left(\sqrt{k_l}\widetilde{r}_l+\sqrt{k_L}\widetilde{r}_L\right)^2}{2\left(\sqrt{k_l}+\sqrt{k_L}\right)}-\frac{\sqrt{k_l}\widetilde{r}_l^2+\sqrt{k_L}\widetilde{r}_L^2}{2}}\int_0^{+\infty}{e^{-\frac{\sqrt{k_l}+\sqrt{k_L}}{2}\left(r-\widehat{r}_{l,L}\right)^2}g(r)dr}\nn
&=N_lN_L e^{-\frac{\sqrt{k_lk_L}\left(\widetilde{r}_l-\widetilde{r}_L\right)^2}{2\left(\sqrt{k_l}+\sqrt{k_L}\right)}}\int_0^{+\infty}{e^{-\frac{\sqrt{k_l}+\sqrt{k_L}}{2}\left(r-\widehat{r}_{l,L}\right)^2}g(r)dr}.\nn
\end{align*}
with $\widetilde{r}_l,k_l$ as defined in (\ref{harmoscD=3}) and $\widehat{r}_{l,L}\!:=\!\frac{\sqrt{k_l}\widetilde{r}_l+\sqrt{k_L}\widetilde{r}_L}{\sqrt{k_l}+\sqrt{k_L}}$. By Taylor expansion of $g(r)$ around $\widehat{r}_{l,L}$,
\begin{align}
\int_0^{+\infty}{e^{-\frac{\sqrt{k_l}+\sqrt{k_L}}{2}\left(r-\widehat{r}_{l,L}\right)^2}g(r)dr}&\simeq\int_{-\infty}^{+\infty}{e^{-\frac{\sqrt{k_l}+\sqrt{k_L}}{2}\left(r-\widehat{r}_{l,L}\right)^2}g(r)dr}\nn
&=\int_{-\infty}^{+\infty}{e^{-\frac{\sqrt{k_l}+\sqrt{k_L}}{2}z^2}\left[\sum_{h=0}^{+\infty}{\frac{g^{(h)}\left(\widehat{r}_{l,L}\right)z^h}{h!} } \right]dr}\nn
&\overset{\left(\ref{esp}\right)}=\sqrt{\frac{2\pi}{\sqrt{k_l}+\sqrt{k_L}}}\sum_{n=0}^{+\infty}{ \frac{g^{(2n)}\left(\widehat{r}_{l,L}\right)}{(2n)!!\left(\sqrt{k_l}+\sqrt{k_L}\right)^n}};\label{appr2}
\end{align}
the equality $\simeq$ holds up to terms vanishing exponentially as $k\to+\infty$, as explained in subsection \ref{Nlproof}. Now, 
\bea
&&N_l=\frac{\sqrt[8]{k_l}}{\sqrt[4]{\pi}}, \qquad k_l
=\frac{2k+4l(l+1)}{2k+3l(l+1)}=1+\frac{l(l+1)}{2k}-\frac{3l^2(l+1)^2}{(2k)^2}+O\left(k^{-3}\right),\nn
&&\sqrt{k_l}=\sqrt{2k}+\frac{3l(l+1)}{2\sqrt{2k}}-\frac{9l^2(l+1)^2}{8(2k)^{\frac{3}{2}}}+O\left(k^{-\frac{5}{2}} \right),\nn
&&\sqrt{k_l}\widetilde{r}_l=\sqrt{2k}+\frac{5l(l+1)}{2\sqrt{2k}}-\frac{21l^2(l+1)^2}{8(2k)^{\frac{3}{2}}}+O\left(k^{-\frac{5}{2}} \right),\nn
&&\frac{\sqrt{k_lk_L}\left(\widetilde{r}_l-\widetilde{r}_L\right)^2}{2\left(\sqrt{k_l}+\sqrt{k_L}\right)}=\frac{\left[l(l+1)-L(L+1)\right]^2}{4\left(2k\right)^{\frac{3}{2}}}+O\left(k^{-\frac{5}{2}}\right)\nn
&&\widehat{r}_{l,L}=1\!+\!\frac{l(l\!+\!1)\!+\!L(L\!+\!1)}{4k}-\frac{\frac{3}{4}l(l\!+\!1)L(L\!+\!1)\!+\!\frac{9}{8}\left[l^2(l\!+\!1)^2\!+\!L^2(L\!+\!1)^2\right]}{4k^2}+O\left(k^{-3}\right).\qquad \label{r_l}
\eea
 But
$$
\sqrt{\frac{2\pi}{\sqrt{k_l}+\sqrt{k_L}}}\frac{\sqrt[8]{k_lk_L}}{\sqrt{\pi}}=\frac{\sqrt{2}}{\sqrt{\sqrt[4]{\frac{k_l}{k_L}}+\sqrt[4]{\frac{k_L}{k_l}}}}=1+O\left(k^{-2}\right),
$$
whence (\ref{general}).
Using relations (\ref{genscp3}),  (\ref{general}), one finds, for example
\bea
\langle\psi_{l-1}^{m+a},g(r)t^a\psi_l^m\rangle = A_l^{a,m}\left(1\!-\!\frac{l^2}{(2k)^{\frac{3}{2}}}\!+\!\cdots\right)\left(g\left(\widehat{r}_{l,l-1}\right)\!+\!\frac{g''\left(\widehat{r}_{l,l-1}\right)}{2\left(\sqrt{k_l}\!+\!\sqrt{k_{l-1}}\right)^2}\!+\!\cdots\!\right),\label{scp+}\\
\langle\psi_{l+1}^{m+a},g(r)t^a\psi_l^m\rangle = B_l^{a,m}\left(1\!-\!\frac{(l\!+\!1)^2}{(2k)^{\frac{3}{2}}}\!+\!\cdots\!\right)\left(g\left(\widehat{r}_{l,l+1}\right)\!+\!\frac{g''\left(\widehat{r}_{l,l+1}\right)}{2\left(\sqrt{k_l}\!+\!\sqrt{k_{l+1}}\right)^2}\!+\!\cdots\!\right.)\label{scp-}
\eea

\subsection{Proof of (\ref{barxpsiD3}) and of proposition \ref{3Dpropo}}
\label{proof3.1}

Using the decomposition $x^a=rt^a$ and the relations (\ref{tY'}) one finds
\bea
\overline{x}^a\psi_{l}^m=\sum_{h,k}{\langle\psi_{h}^k,x^a\psi_{l}^m\rangle\psi_{h}^k}=\left\{\!\!
\ba{ll}
\langle \psi_{l-1}^{m+a},r t^a\psi_{l}^{m}\rangle\psi_{l-1}^{m+a}+
\langle \psi_{l+1}^{m+a},r t^a\psi_{l}^{m}\rangle\psi_{l+1}^{m+a
}&\mbox{ if }l<\Lambda,\\[8pt]
\langle \psi_{\Lambda-1}^{m+a},r t^a\psi_{\Lambda}^{m}\rangle\psi_{\Lambda-1}^{m+a}&\mbox{ if }
l=\Lambda,\\[8pt]
0&\mbox{otherwise.}
\ea
\right.   \label{barxpsiD3'}
\eea
From (\ref{scp+}) and (\ref{scp-}) one has
\bea
\ba{l}
\hspace{2.5cm}\langle \psi_{l-1}^{m+a},r t^a\psi_{l}^{m}\rangle=c_l A_{l}^{a,m},\hspace{1cm}\langle \psi_{l+1}^{m+a},r t^a\psi_{l}^{m}\rangle=c_{l+1}B_l^{a,m},\\[8pt]
\mbox{where, up to }O\left(k^{-\frac{3}{2}} \right),\quad c_l= 1+\frac{l^2}{2k}=\sqrt{1+\frac{l^2}{k}}\quad 1\le l\le \Lambda,\qquad c_0=c_{\Lambda+1}=0.
\ea    \label{factorization}
\eea
$c_l$ is an integral over the $r$ variable, while $A_{l}^{a,m}$ is an integral over the angle variables;
replacing in  (\ref{barxpsiD3'}) we find  (\ref{barxpsiD3}). The  factorizations (\ref{factorization}) are manifestations
 of the Wigner-Eckart theorem. 

Using (\ref{barxpsiD3}) one can calculate the action of the commutator on $\psi_{l}^{m}$.
For all $l<\Lambda$ and $m$ with $|m|\le l$, one has
$$
[\overline{x}^+,\overline{x}^-]\psi_{l}^{m}=\left[(c_l)^2-(c_{l+1})^2\right]\frac{L_0}{2l+1}\psi_{l}^{m},
\qquad\qquad [\overline{x}^{0},\overline{x}^\pm]\psi_{l}^{m}=\left[(c_l)^2-(c_{l+1})^2\right]\frac{\pm L_\pm}{2l+1}\psi_{l}^{m}
$$
Since
$$
\left[(c_l)^2-(c_{l+1})^2\right]=-\frac{2l\!+\!1}{k}+O\left(\frac{1}{k^{\frac{3}{2}}} \right)
\mbox{ if }\: l<\Lambda\mbox{ and }c_{\Lambda}^2= 1\!+\!\frac{\Lambda^2}{k}+O\left(\frac{1}{k^{\frac{3}{2}}} \right)
$$
we find at leading order in $1/\sqrt{k}$                 
\be
[\overline{x}^+,\overline{x}^-]=-\frac{L_0}{k}+C\widetilde{P}_{\Lambda}L_0,\qquad\quad [\overline{x}^{0},\overline{x}^\pm]=\mp\frac{L_\pm}{k}
\pm C
\widetilde{P}_{\Lambda}L_\pm,\qquad  C:=\frac 1k+\frac{1+\frac{\Lambda^2}{k}}{2\Lambda+1}.     \label{cr3D}
\ee
Going back to the hermitean coordinates
 $\overline{x}\equiv \overline{x}^1=\frac{\overline{x}^++\overline{x}^-}{\sqrt{2}}$ and $\overline{y}\equiv \overline{x}^2=\frac{\overline{x}^+-\overline{x}^-}{\sqrt{2}i}$ one finds
$$
[\overline{x}^i,\overline{x}^j]=i\varepsilon^{ijh}\left(-\frac{I}{k}+C\widetilde{P}_{\Lambda}\right)\overline{L}_h
$$
In this model the commutator is invariant under parity because the components of the angular momentum are pseudo-vectors and they do not change sign under parity, while $P_{\overline{E}}$ is a projector, then a scalar.

Similarly, one has
\be
\mathcal{R}^2\psi_l^m=\left\{
\ba{lr}
\left[1+\frac{1}{k}(l^2+l+1)+O\left(\frac{1}{k^{\frac{3}{2}}}\right)\right]\psi_{l}^{m}&\mbox{ if }l<\Lambda\\
\left[\left(1+\frac{\Lambda^2}{k}\right)\left(\frac{\Lambda}{2\Lambda+1}\right)+O\left(\frac{1}{k^{\frac{3}{2}}}\right)\right]\psi_{\Lambda}^{m}&\mbox{if }l=\Lambda
\ea
\right. \label{r23d}
\ee
or equivalently,  at leading order in $1/\sqrt{k}$,
\be
\mathcal{R}^2:= \sum_{l=0}^{\Lambda-1}\left[1
+\frac{1}{k}(l^2+l+1)\right]\widetilde P_l+\left(1+\frac{\Lambda^2}{k}\right)\left(\frac{\Lambda}{2\Lambda+1}\right)\widetilde P_{\Lambda}. 
\ee
We see that the square distance from the origin is not central and equal to $1$ on all the representation (as in the standard quantization on the unit sphere), but its eigenvalues go all to $1$ as $k\rightarrow +\infty$, except when $l=\Lambda$.
The last relation and (\ref{cr3D}) are exact if (\ref{defLD=3}-\ref{defc_l})  
are adopted as exact definitions.

In order to prove that $\overline{x}\cdot\overline{L}=0$, we must recall equation (\ref{trasfcoord}) and since (by definition) $L^j=\frac{1}{i}\varepsilon^{jkh}x^k\partial_h$ one has $x\cdot L=\frac{1}{i}\varepsilon^{jkh}x^jx^k\partial_h$, but $\varepsilon^{jkh}$ is anti-symmetric and $x^jx^k$ is symmetric, then $x\cdot L=0$. If one projects the previous operators on the Hilbert space $\mathcal{H}_{\Lambda}$, since when $L^j$ acts on $\psi_{l}^m$ it preserves the square of the angular moment (because $L^j$ doesn't increase the value of $l$), that is $P_{\overline{E}}L^j=L^jP_{\overline{E}}$ then one has $P_{\overline{E}}x^jL^jP_{\overline{E}}=P_{\overline{E}}x^jP_{\overline{E}}L^j$; in conclusion, the condition $x\cdot L=0$ imples $\overline{x}\cdot\overline{L}=0$. One can also verify this condition in this way, from $L_{\pm}=\frac{L_x\pm iL_y}{\sqrt{2}}$, then one has
\bea
&&\overline{x}\cdot \overline{L}=\overline{x}^iL^j\eta_{ij}=\overline{x}^aL^b\widetilde{\eta}_{ab}=\left(\overline{x}^+L_-+\overline{x}^-L_++\overline{x}^0L_0\right)\psi_l^m\nn
&&+c_l\left(\frac{1}{2}\sqrt{l(l+1)-m(m-1)}A_l^{+,m-1}+\frac{1}{2}\sqrt{l(l+1)-m(m+1)}A_l^{-,m+1}+mA_l^{0,m}\right)\psi_{l-1,m}\nn
&&+c_{l+1}\left(\frac{1}{2}\sqrt{l(l\!+\!1)\!-\!m(m\!-\!1)}B_l^{+,m-1}+\frac{1}{2}\sqrt{l(l\!+\!1)\!-\!m(m\!+\!1)}B_l^{-,m+1}\!+\! mB_l^{0,m}\right)\psi_{l+1,m}\nonumber
\eea
and using (\ref{ClebschAB}) one obtains $\overline{x}\cdot \overline{L}=0$.
\hfill\qedsymbol

\subsection{Action and commutators  of the $\overline{\partial}_a$}

In order to calculate the action of the derivation operators on the Hilbert space $\mathcal{H}_{\Lambda}$, we first note that 
$Y_l^mr^l$ is a homogeneous polynomial of degree $l$ in the $x^a$ variables. This implies $\partial_a\left(Y_l^mr^l\right)=C_l^{a,m}\left(Y_{l-1}^{m-a}r^{l-1}\right)$, with some coefficients $C_l^{a,m}$. From the identities
$$
\left[\partial_a,\partial_b\right]\left(Y_l^m r^l\right)=0, \qquad
\left(\left[\partial_a,x^b\right]-\delta_a^b\right)\left(Y_l^mr^l\right)=0
$$
one finds
\bea
&&C_l^{b,m}C_{l-1}^{a,m-b}-C_l^{a,m}C_{l-1}^{b,m-a}=0,\nn
&&A_l^{b,m}C_{l-1}^{a,m+b}+2A_l^{b,m}A_{l-1}^{-a,m+b}-C_l^{a,m}A_{l-1}^{b,m-a}=0,\label{AS}\\
&&2A_l^{b,m}A_{l}^{a,m+b-a}+A_{l+1}^{-b,m+b}C_{l+1}^{a,m+b}-C_l^{a,m}A_{l}^{-b,m-a+b}=\delta_a^b.\nonumber
\nonumber
\eea
By explicit calculations, one can conclude that $C^{a,m}_l=(2l+1) A^{-a,m}_l$.
On the other hand, using 
\bea
\partial_a\frac{f_l(r)}{r^{l+1}}&=&\frac{\partial}{\partial x^a}\frac{f_l\left(\sqrt{x^bx^{-b}}\right)}{\left(x^bx^{-b}\right)^{\frac{l+1}{2}}}=-\frac{l+1}{2}2x^{-a}\frac{f_l\left(\sqrt{x^bx^{-b}}\right)}{\left({x^bx^{-b}}\right)^{\frac{l+3}{2}}}+\frac{f'_l\left(\sqrt{x^bx^{-b}}\right)}{\left(x^bx^{-b}\right)^{\frac{l+1}{2}}}\frac{2x^{-a}}{2\sqrt{x^bx^{-b}}}\nn
&=&\left[\frac{f'_l(r)}{r^{l+1}}-\frac{l+1}{r^{l+2}}f_l(r)\right]t^{-a}, \nonumber
\eea
and the results of section \ref{scp3D} we find
\bea
\partial_a\psi_l^m\left(r,\theta,\varphi\right)&=&\partial_a\left(\frac{f_l(r)}{r^{l+1}} \right)r^lY_l^m\left(\theta,\varphi\right)+\frac{f_l(r)}{r^2}C_l^{a,m}Y_{l-1}^{m-a}\left(\theta,\varphi\right)\nn
&=&\left[\frac{f'_l(r)}{r}-\frac{l+1}{r^2}f_l(r)\right]t^{-a}Y_l^m\left(\theta,\varphi\right)+\frac{f_l(r)}{r^2}C_l^{a,m}Y_{l-1}^{m-a}\left(\theta,\varphi\right)\nn
&=&\left[\frac{f'_l(r)}{r}-(l\!+\!1)\frac{f_l(r)}{r^2}\right]B^{-a,m}_lY^{m-a}_{l+1}+\left[\frac{f'_l(r)}{r}+l\frac{f_l(r)}{r^2}\right] A^{-a,m}_lY_{l-1}^{m-a}.\nonumber
\eea
Hence
\bea
\overline{\partial}_a\psi_l^m = \psi_{l-1}^{m-a}\left[M_l+lJ_l\right]A^{-a,m}_l-\psi_{l+1}^{m-a}\left[M_{l+1}+(l\!+\!1)J_{l+1}\right]B^{-a,m}_l,
\eea
where
\bea
&&J_l:=\int_{0}^{+\infty}{\frac{f_l(r)f_{l-1}(r)}{r}dr}=1+\frac{1}{\sqrt{8k}}-\frac{l^2}{2k}+O\left(\frac{1}{k^{\frac{3}{2}}}\right),\\
&&M_l:=\int_{0}^{+\infty}{f_{l-1}(r)f'_l(r)dr}=-\frac{l}{\sqrt{2k}}+\frac{18l^3}{8k\sqrt{2k}}+O\left(\frac{1}{k^{\frac{3}{2}}}\right);
\label{JlMl}
\eea
$J_l,M_l$ have been evaluated using (\ref{general}).
The commutator $\left[\overline{\partial}_a,\overline{\partial}_b\right]$ on $\psi_l^m$ is
\bea
&\left[\overline{\partial}_a,\overline{\partial}_b\right]\psi_l^m=&\left[\left(J_{l+1}\right)^2\frac{(l+1)^2}{2l+1}-\left(J_l\right)^2\frac{l^2}{2l+1}+2(l+1) J_{l+1}M_{l+1}-2lJ_lM_l\right.\nn
&&\left.\hspace{3.5cm}+\left(M_{l+1}\right)^2\frac{1}{2l+1}-\left(M_l\right)^2\frac{1}{2l+1}\right]\alpha_{a,b} \overline{L}_{-a-b}\psi_l^m\nn
\nonumber
\eea
with
$$
\alpha_{a,b}=-\alpha_{b,a}=
\left\{
\begin{array}{ll}
0&\mbox{ if }a=b\\
-1&\mbox{ if }a=0,b=+1\\
1&\mbox{ if }a=0,b=-1\\
1&\mbox{ if }a=-1,b=+1\\
\end{array}
\right.
$$

\subsection{Proof of proposition \ref{realso(4)'} and other results of subsection \ref{realso(4)}}
\label{cal}

Let $\chi_i:=i\varepsilon^{ijk}X_jL_k$. Using (\ref{CRXL})  we 
easily find
\bea
L^2X_i=X_iL^2+2(X_i+\chi_i),\qquad L^2\chi_i=(\chi_i+2X_i)L^2. 
\label{utile2}
\eea
This suggests to look for  ``eigenvectors" of $L^2$ (and therefore of $\lambda=[\sqrt{4L^2+1}-1]/2$), \ 
$L^2\vartheta_i=\vartheta_i\,\nu(\lambda)$, \   in the form $\vartheta_i=X_ia(\lambda)+\chi_ib(\lambda)$. The compatibility condition
is a second degree equation with the ``right eigenvalue" $\nu(\lambda)$ as the unknown.
Up to $\lambda$-dependent factors the solutions are
\bea
\ba{ll}
\vartheta_i^-=\left[X_i\lambda-\chi_i \right]  (2\lambda\!-\!1) (\lambda\!-\!1), \qquad &\nu^-:=(\lambda\!-\!1)\lambda,\\[8pt]
\vartheta_i^+=\left[X_i(\lambda\!+\!1)+\chi_i \right]\, \lambda(2\lambda\!+\!3) \qquad &\nu^+:=(\lambda\!+\!1)(\lambda\!+\!2)
\ea
\label{L^2eigen}
\eea
(here we have chosen the coefficients $a(\lambda),b(\lambda)$ so that $\vartheta_i^\pm{}^\dagger=\vartheta_i^\mp$); this implies 
\be
\lambda\,\vartheta_i^\pm=\vartheta_i^\pm \,(\lambda\pm 1).
\ee
Therefore \ $\vartheta_a^\pm\vert l,m\rangle\propto 
\vert l\pm 1,m+a\rangle$.
Inverting (\ref{L^2eigen}) one can express the $X_a$ (as well as the $\chi_a$)
as linear combinations of $\vartheta_a^\pm$ with $\lambda$-dependent 
coefficients; hence $\alpha_{l,j}^{a,m}=0$ unless $j=l\pm 1$, and the Ansatz
(\ref{u4}), as anticipated.

\medskip
On the other hand, using (\ref{AB}-\ref{Aproperties}) and the equalities
$$
\begin{array}{lcl}
A_{l}^{-,m}B_{l-1}^{+,m+1}-A_{l}^{+,m}B_{l-1}^{-,m+1}=\frac{m}{2l+1},& & A_{l}^{\pm,m}B_{l-1}^{0,m\pm1}-A_{l}^{0,m}B_{l-1}^{\pm,m}=\frac{\pm\gamma_{l}^{\pm,m}}{2l+1},\\
\\
d_l^2-d_{l+1}^2=2l+1,& &A_l^{a,m}\gamma_{l-1}^{a,m+a}=A_l^{a,m+a}\gamma_l^{a,m},\\
\\
B_l^{a,m}\gamma_{l+1}^{a,m+a}=B_l^{a,m+a}\gamma_l^{a,m}, & &A_l^{0,m}\gamma_{l-1}^{\pm,m}-A_l^{0,m\pm 1}\gamma_l^{\pm,m}=\mp A_l^{\pm,m},\\
\\
B_l^{0,m}\gamma_{l+1}^{\pm,m}-B_l^{0,m\pm 1}\gamma_l^{\pm,m}=\mp B_l^{\pm,m},& &A_l^{\pm,m}\gamma_{l-1}^{\mp,m\pm 1}-A_l^{\pm,m\mp 1}\gamma_l^{\mp,m}=\mp A_l^{0,m},\\
\\
B_l^{\pm,m}\gamma_{l+1}^{\mp,m\pm 1}-B_l^{0,m\mp 1}\gamma_l^{\mp,m}=\mp B_l^{0,m},& & \\
\end{array}
$$
we obtain, for example,
\bea
&\left[X_{+},X_{-}\right]\left\vert l,m\rangle\right.=&d_ld_{l-1}\left(A_l^{-,m}A_{l-1}^{+,m-1}-A_l^{+,m}A_{l-1}^{-,m+1}\right)\left\vert l-2,m\rangle\right.+\nn
&&d_l^2\left(A_l^{-,m}B_{l-1}^{+,m-1}-A_l^{+,m}B_{l-1}^{-,m+1}\right)\left\vert l,m\rangle\right.+\nn
&&d_{l+1}^2\left(B_l^{-,m}A_{l+1}^{+,m-1}-B_l^{+,m}A_{l+1}^{-,m+1}\right)\left\vert l,m\rangle\right.+\nn
&&d_{l+1}d_{l+2}\left(B_l^{-,m}B_{l+1}^{+,m-1}-B_l^{+,m}B_{l+1}^{-,m+1}\right)\left\vert l+2,m\rangle\right.=\nn
&&\left(d_l^2-d_{l+1}^2\right)\left(A_l^{-,m}B_{l-1}^{+,m-1}-A_l^{+,m}B_{l-1}^{-,m+1}\right)\left\vert l,m\rangle\right.=\nn
&&2l+1\frac{m}{2l+1}\left\vert l,m\rangle\right.=m \left\vert l,m\rangle\right.=L_0 \left\vert l,m\rangle\right. ,\nn
&\left[X_{+},X_{0}\right]\left\vert l,m\rangle\right.=&d_ld_{l-1}\left(A_l^{0,m}A_{l-1}^{+,m}-A_l^{+,m}A_{l-1}^{0,m+1}\right)\left\vert l-2,m+1\rangle\right.+\nn
&&d_l^2\left(A_l^{0,m}B_{l-1}^{+,m}-A_l^{+,m}B_{l-1}^{0,m+1}\right)\left\vert l,m+1\rangle\right.+\nn
&&d_{l+1}^2\left(B_l^{0,m}A_{l+1}^{+,m}-B_l^{+,m}A_{l+1}^{0,m+1}\right)\left\vert l,m+1\rangle\right.+\nn
&&d_{l+1}d_{l+2}\left(B_l^{0,m}B_{l+1}^{+,m}-B_l^{+,m}B_{l+1}^{0,m+1}\right)\left\vert l+2,m+1\rangle\right.=\nn
&&\left(d_l^2-d_{l+1}^2\right)\left(A_l^{0,m}B_{l-1}^{+,m}-A_l^{+,m}B_{l-1}^{0,m+1}\right)\left\vert l,m+1\rangle\right.=\nn
&&2l+1\frac{-\gamma_l^{+,m}}{2l+1}\left\vert l,m+1\rangle\right.=-\gamma_l^{+,m} \left\vert l,m\rangle\right.=-L_+ 
\left\vert l,m\rangle\right. .\nn
\nonumber
\eea
Since the Ansatz (\ref{u4}) differs from (\ref{tY'}) only by the $l$-dependent
coefficients $d_l$, and $l$ is not changed by the action of the $L_b$, the fact that 
(\ref{basic_so3'}) holds for $v^a=t^a$ guarantees that it holds also for  $v^a=X_a$ , i.e. proves the relations of the form \ $\left[L_a,X_b\right]=f^c_{ab}X_c$ \ in (\ref{u2}). 

In addition, because of
\be
A_l^{+,m-1}\gamma_l^{-,m}+A_l^{-,m-1}\gamma_l^{+,m}+A_l^{0,m}m=B_l^{+,m-1}\gamma_l^{-,m}+B_l^{-,m-1}\gamma_l^{+,m}+B_l^{0,m} m=0,
\ee
\be
A_l^{+,m}\gamma_{l-1}^{-,m+1}+A_l^{-,m}\gamma_{l-1}^{+,m-1}+A_l^{0,m}m=B_l^{+,m}\gamma_{l+1}^{-,m+1}+B_l^{-,m}\gamma_{l+1}^{+,m-1}+B_l^{0,m}m =0
\ee
and
\be
A_l^{0,m}B_{l-1}^{0,m}+A_l^{-,m}B_{l-1}^{+,m-1}+A_l^{+,m}B_{l-1}^{-,m+1}=\frac{l}{2l+1}
\ee
we obtain
\be
X^2\left\vert l,m \rangle\right.=\left[d_{l+1}^2+\left(d_l^2-d_{l+1}^2\right)\frac{l}{2l+1}\right]\left\vert l,m \rangle\right.=\left[\Lambda(\Lambda+2)-l(l+1)\right]\left\vert l,m \rangle\right.
\ee
whence we can easily derive (\ref{u3}).

\medskip
To prove (\ref{gO(3)}) we first note that, using the basic property  $\Gamma(z+1)=z \Gamma(z)$,
\bea
\prod_{h=0}^{l-1}(\Lambda\!+\!l\!-\!2h)=2^{l}\prod_{h=0}^{l-1}\left(\frac{\Lambda\!+\!l}2\!-\!h\right)
=2^{l}\frac{\Gamma\!\left(\frac {\Lambda\!+\!l}2\!+\!1\right)}
{\Gamma\!\left(\frac {\Lambda\!-\!l}2\!+\!1\right)}=:h_1(l)\nn
\prod_{h=0}^l(\Lambda\!+\!l\!+\!1\!-\!2h)=2^{l+1}\prod_{h=0}^{l}\left(\frac{\Lambda\!+\!l\!+\!1}2\!-\!h\right) =2^{l+1}\frac{\Gamma\!\left(\frac {\Lambda\!+\!l\!+\!1}2\!+\!1\right)}
{\Gamma\!\left(\frac {\Lambda\!-\!l\!+\!1}2\right)}=:h_2(l).\nonumber
\eea
Hence $h_1(l)/h_2(l)$ gives the first ratio under the square root in (\ref{gO(3)'}), implying
\bea
\frac{h_1(l)}{h_2(l)}\,\frac{h_1(l\!-\!1)}{h_2(l\!-\!1)}=\frac{\Gamma\!\left(\frac {\Lambda\!+\!l\!+\!1}2\right)\Gamma\!\left(\frac {\Lambda\!-\!l\!+\!1}2\right)}
{4\: \Gamma\!\left(\frac {l\!+\!\Lambda\!+\!1}2\!+\!1\right)\Gamma\!\left(\frac {\Lambda\!-\!l\!+\!1}2\!+\!1\right)}
=\frac{1}{(\Lambda\!+\!1\!+\!l)(\Lambda\!+\!1\!-\!l)}.
\eea
On the other hand, it is \ $1+\frac{l^2}k=\frac 1k (l\!+\!i\sqrt{k})(l\!-\!i\sqrt{k})$. \ Setting \
$
h_\pm(l)\!:=\! \sqrt{2}\frac{\Gamma\left(\frac l2\!+\!1\!\pm\!\frac{i\sqrt{k}}2\right)}
{\Gamma\left(\frac {l\!+\!1}2\!\pm\!\frac{i\sqrt{k}}2\right)}
$
\ we find
\bea
h_\pm(l)h_\pm(l-1)=2\frac{\Gamma\left(\frac l2\!+\!1\!\pm\!\frac{i\sqrt{k}}2\right)}
{\Gamma\left(\frac {l\!+\!1}2\!\pm\!\frac{i\sqrt{k}}2\right)}
\frac{\Gamma\left(\frac {l\!+\!1}2\!\pm\!\frac{i\sqrt{k}}2\right)}
{\Gamma\left(\frac {l}2\!\pm\!\frac{i\sqrt{k}}2\right)}=l\!\pm\!i\sqrt{k}
\quad\Rightarrow\quad 1+\frac{l^2}k=  f(l)f(l\!-\!1)
\eea
where $f(l):=h_+(l)h_-(l)/\sqrt{k}$. 
Therefore $g(l)=\sqrt{h_1(l)f(l)/h_2(l)}$, i.e. (\ref{gO(3)}), solves (\ref{gcond}).

\subsection{Shifting the lower extreme of integration over $r$}
\label{Nlproof}

Here we justify the approximation used in (\ref{appr1}) and (\ref{appr2}):
\be
\int_0^{+\infty}\!\! e^{-(r-\widetilde{r}_l)^2\sqrt{k_l}}g(r) dr\simeq \int_{-\infty}^{+\infty}\!\! e^{-(r-\widetilde{r}_l)^2\sqrt{k_l}} g(r)  dr.
\ee
We first consider $g(r)\equiv 1$ and estimate for $b>0$ and $a\to+\infty$ the following difference:
\bea
 \left(\int_{-\infty}^{+\infty}\!\!-\int_0^{+\infty}\right) e^{-a(r-b)^2}dr &=& \int_0^{+\infty}\!\!\!\! e^{-a(r+b)^2}dr
=\int_{b\sqrt{a}}^{+\infty}\!\!\!\! e^{-z^2}\frac{dz}{\sqrt{a}}=\frac{\sqrt{\pi}}{2\sqrt{a}}\:\mbox{erfc}\left(b\sqrt{a}\right)\nn
&=&\frac{e^{-a b^2}}{2ba}\left[1\!-\!\frac 1{2ab^2}+\cdots\right].                       \label{error}
\eea
Here we have used: the changes  of integration variables $r\mapsto -r$,  $r\mapsto z=\sqrt{a}(y+b)$ in the first, second equalities;
 the definition \ $\mbox{erfc}(x)=\frac2 {\sqrt{\pi}}\int_{x}^{+\infty}\!\!\!\! e^{-z^2}dz$ \
and the $x\to\infty$ asymptotic expansion \ $\mbox{erfc}(x)=\frac{ e^{-x^2}}{x\sqrt{\pi}}\left[1\!-\!\frac 1{2x^2}\!+\!\cdots\right]$ \
of the complementary error function in the third and fourth.  We can apply the above formula to the integral (\ref{appr1}) setting $a=\sqrt{2k}\!+\!\cdots$, $b=1\!+\!\cdots$; consequently,
 the error made shifting from 0 to $-\infty$ the lower  extreme in integrating over $r$ is of the order $1/\left(2\sqrt{2k}e^{\sqrt{2k}}\right)$, which has zero asymptotic expansion in $1/\sqrt{2k}$, hence does not contribute to the expansion of the integral: here are the meaning
and the justification of the symbol $\simeq$. One obtains a similar
result after a suitable number of integration by parts also if $g(r)$ is polynomial (or more generally analytic);  this justifies (\ref{appr2}).

\subsection{Proof of proposition \ref{propoD3} and other results of section \ref{3Dconverge}}
\label{convergD3}

We first show that the operators $\overline{x}^a$ converge strongly to $t^{a}$.
From  (\ref{barxpsiD3}-\ref{tY'})
we find
\bea
(\overline{x}^a\!-\!t^a)\phi=(\overline{x}^a\!-\!t^a)\sum_{l\in\mathbb{N}_0}\sum_{m=-l}^{l}\phi_l^m Y_l^m=\sum_{l=0}^{\infty}\sum_{m=-l}^{l}\phi_l^m\left[(c_l\!-\!1)A_l^{a,m}Y_{l-1}^{m+a}+(c_{l+1}\!-\!1)B_l^{a,m}Y_{l+1}^{m+a}\right],
\nonumber
\eea
with  $c_l\equiv 0$ for $l\!>\!\Lambda$. This implies
\bea
\Vert(\overline{x}^a\!-\!t^a)\phi\Vert^2=
\sum_{l=0}^{\infty}\sum_{m=-l}^{l}\left\{|\phi_l^m|^2\left[(c_l\!-\!1)^2 \left|A_l^{a,m}\right|^2+(c_{l+1}\!-\!1)^2\left|B_l^{a,m}\right|^2\right]\right.\nn
\left. +A_{l+2}^{a,m}B_l^{a,m}(c_{l+2}\!-\!1)(c_{l+1}\!-\!1)\left[\overline{\phi_{l+2}^m}\phi_l^m\!+\!\phi_{l+2}^m\overline{\phi_l^m}\right]\right\}\nn
\le \sum_{l=0}^{\infty}\sum_{m=-l}^{l}\left\{\frac{|\phi_l^m|^2}2\left[(c_l\!-\!1)^2 +(c_{l+1}\!-\!1)^2\right]
 +\frac 12 (c_{l+2}\!-\!1)(c_{l+1}\!-\!1)\left[\left\vert\phi_{l+2}^m\right\vert^2 \!+\! \vert\phi_l^m\vert^2\right]\right\}\nn
\le\sum_{l=0}^{\Lambda}\sum_{m=-l}^{l}\left\vert\phi_l^m\right\vert ^2\frac{\Lambda^4}{2k^2}+\sum_{l\geq \Lambda+1}\sum_{m=-l}^{l}2\left\vert\phi_l^m\right\vert^2\leq\frac{\Lambda^4}{2k^2}\left\|\phi\right\|^2+2\sum_{l\geq \Lambda+1}\sum_{m=-l}^{l}\left\vert\phi_l^m\right\vert^2;
\label{ineqs}
\eea
here we have used the inequalities
\be
\left\vert A_l^{a,m}\right\vert,\left\vert B_l^{a,m}\right\vert\leq \frac{1}{\sqrt{2}},\quad\qquad 
(c_l-1)\leq \frac{l^2}{2k}\leq\frac{\Lambda^2}{2k }\quad\mbox{if } l\le \Lambda.\label{stime}
\ee
(the second one follows from $\sqrt{1\!+\!\varepsilon}\!-\!1\!<\! \varepsilon/2$
for $\varepsilon\!>\!0$).  By  (\ref{consistency}) 
 the right-hand side of (\ref{ineqs})  goes to zero as $\Lambda\to\infty$,
as claimed.

Since $\overline{x}^a$ annihilates $\Hi_\Lambda^\perp$, 
$\overline{x}^a$  {\it cannot} converge  to $t^a$ {\it in operator norm}.
In fact,  the square norm 
e.g. of \ $(t^\pm\!-\!\overline{x}^\pm)Y_{\Lambda+1}^{\pm(\Lambda+1)}\!=\!  B_{\Lambda+1}^{\pm,\pm(\Lambda+1)}Y_{\Lambda+2}^{\pm(\Lambda+2)}$ \ is \ $\left|B_{\Lambda+1}^{\pm,\pm(\Lambda+1)}\right|^2=\frac{2\Lambda\!+\!4}{2(2\Lambda\!+\!5)}\ge 3/7$, 
implying $\Vert\overline{x}^\pm\!-\!t^\pm\Vert\ge\sqrt{3/7}$ for all $\Lambda$. Similarly,  the square norm of $(t^0\!-\!\overline{x}^0)Y_{\Lambda+1}^{0}\!=\! A_{\Lambda+1}^{0,0}Y_{\Lambda}^{0}\!+\! B_{\Lambda+1}^{0,0}Y_{\Lambda+2}^{0}$ is $\left|A_{\Lambda+1}^{0,0}\right|^2\!+\! \left|B_{\Lambda+1}^{0,0}\right|^2=\frac{(\Lambda\!+\!2)^2\!+\!(\Lambda\!+\!1)^2}{(2\Lambda\!+\!3)(2\Lambda\!+\!5)}\ge 1/3$, implying $\Vert\overline{x}^0\!-\!t^0\Vert\ge \sqrt{1/3}$ for all $\Lambda$.

\medskip
We now prove \ref{propoD3}. Since
\bea
(f -\hat f_\Lambda)\phi=\sum_{l=0}^{\Lambda}\sum_{|m|\leq l}Y_l^m \chi_l^m
+\sum_{l>\Lambda}\sum_{|m|\leq l}Y_l^m (f\phi)_l^m , 
\eea
where \ $\chi_l^m:=(f\phi)_l^m-(\hat f_\Lambda \phi)_l^m$,  $(f\phi)_l^m= \langle Y_l^m,f\phi\rangle$, $ \left(\hat f_\Lambda \phi\right)_l^m=\langle Y_l^m,\hat{f}_\Lambda\phi\rangle$], \ we find
\bea
 \Vert(f -\hat f_\Lambda)\phi\Vert^2=\sum_{l=0}^{\Lambda}\sum_{|m|\leq l}|\chi_l^m|^2
+\sum_{l>\Lambda}\sum_{|m|\leq l}|(f\phi)_l^m|^2.        \label{normffh}
\eea
As the second sum goes to zero as $\Lambda\to\infty$, it remains to show that the first sum
does as well. We find
\bea
\chi_l^m=\left\langle Y_l^m,\left( f-\hat{f}_{\Lambda}\right)\phi\right\rangle=\left\langle Y_l^m,\sum_{j=0}^{2\Lambda}\sum_{|s|\leq j}f_j^s\left(Y_j^s-\widehat{Y}_j^s\right)
\phi\right\rangle \nn 
=\sum_{j=0}^{2\Lambda}\sum_{|s|\leq j}f_j^s \:\: R_j^s {\sum}'
 \left\langle Y_l^m,\pm\left( t^{a_1}\cdots t^{a_j}-\overline{x}^{a_1}\cdots \overline{x}^{a_j}\right)\phi\right\rangle \nn 
=\sum_{j=0}^{2\Lambda}\sum_{|s|\leq j}f_j^s 
\:\: R_j^s {\sum}' \sum_{l'=0}^{\infty}\sum_{m'=-l'}^{l'}\phi_{l'}^{m'}
 \left\langle Y_l^m,\pm\left( t^{a_1}\cdots t^{a_j}-\overline{x}^{a_1}\cdots \overline{x}^{a_j}\right)Y_{l'}^{m'}\right\rangle,           \label{chi}
\eea
where $\sum'$ is the sum (\ref{differenceYlm}) of  $n\le j^{j-s}$ 
(not necessarily distinct) monomials in $\overline{x}^+,\overline{x}^0,\overline{x}^{-1}$ with coefficients $\pm 1$. 
The computation when $j=2$ is instructive for the generic situation: 
\bea
\left(\overline{x}^{a}  \overline{x}^{b}\!-\!t^{a} t^{b}\right)Y_l^m=
(c_lc_{l-1}\!-\!1)
A_l^{b,m}A_{l-1}^{a,m+b}Y_{l-2}^{m+a+b}
+(c_{l+1}c_{l+2}\!-\!1)B_l^{b,m}B_{l+1}^{a,m+b}Y_{l+2}^{m+a+b}\nn +\left[\left(c_l^2\!-\!1\right)A_l^{b,m}B_{l-1}^{a,m+b}\!+\!
\left(c_{l+1}^2\!-\!1\right)B_l^{b,m}A_{l+1}^{a,m+b}
\right]Y_l^{m+a+b}.
\nonumber
\eea
More generally
\be
\left(t^{a_1}\cdots t^{a_j}\!-\!\overline{x}^{a_1}\cdots  \overline{x}^{a_j}\right)Y_{l'}^{m'}=\sum_{h=0}^{j}A^{a_1,\cdots,a_j}_{l',m',h}Y_{l'-j+2h}^{m'+s},\label{t-x}
\ee
where $A^{a_1,\cdots,a_j}_{l',m',h}$ is a sum of at most $j\choose {\left[\frac{j+1}{2}\right]}$ terms of the form
\be
\prod_{h=0}^{j}A_{l'_h}^{a'_h,m'_h}\left(1-\prod_{h'=0}^{j}c_{l'_{h'}}\right),
\label{fattori}
\ee
where $a'_h=a_h$ when the factor comes from a coefficient $A$, while $a'_h=-a_h$ when it comes from a $B$.
By (\ref{stime}) the factors in (\ref{fattori}) satisfy the inequalities
\bea
&& \left\vert\prod_{h=0}^{j}\! A_{l'_h}^{a'_h,m'_h} \right\vert\leq\frac{1}{2^{\frac{j}{2}}},\quad\left\vert 1-\!\prod_{h'=0}^{j}\! c_{l'_{h'}}\!\right\vert\leq \left(\!1+\frac{\Lambda^2}{k}\right)^{\frac{j}{2}}\!-1\leq e^{\frac{j\Lambda^2}{2k}}-1 \nn[8pt]
&& \Rightarrow \qquad\qquad
\left\vert\! A^{a_1,\cdots,a_j}_{l,m,h} \!\right\vert\leq
{{j}\choose{\left[\frac{j+1}{2}\right]}}\frac{e^{\frac{j\Lambda^2}{2k}}-1}{2^{\frac{j}{2}}}.                         \label{stime2}
\eea
If we replace (\ref{t-x}) into (\ref{chi}) and use \
$\left\langle Y_l^m,Y_{l'}^{m'}\right\rangle=\delta_{l}^{l'}\delta_m^{m'}$ \
we obtain
\bea
&\chi_l^m&=\sum_{j=0}^{2\Lambda}\sum_{|s|\leq j}f_j^s 
\:\: R_j^s {\sum}' 
\sum_{l'=0}^{\infty}\sum_{m'=-l'}^{l'}\phi_{l'}^{m'}
\sum_{h=0}^{j} \pm \left\langle Y_l^m,A^{a_1,\cdots,a_j}_{l',m',h}Y_{l'-j+2h}^{m'+s}
\right\rangle\nn
&&=\sum_{j=0}^{2\Lambda}\sum_{|s|\leq j}f_j^s 
\:\: R_j^s \:\: {\sum}'
\sum_{l'=0}^{\infty}\sum_{m'=-l'}^{l'}\phi_{l'}^{m'}
 \sum_{h=0}^{j}\pm A^{a_1,\cdots,a_j}_{l',m',h}\delta_{l}^{l'-j+2h}\delta_{m}^{m'+s}\nn
&&=\sum_{j=0}^{2\Lambda}\sum_{|s|\leq j}f_j^s 
\:\: R_j^s \:\: {\sum}'
\sum_{0\leq h\leq\min\left\{j,\frac{l+j-|m-s|}{2}\right\}}\pm\phi_{l+j-2h}^{m-s}A^{a_1,\cdots,a_j}_{l+j-2h,m-s,h}\nonumber
\eea
and, by (\ref{chi}-\ref{stime2}), 
$$
\left\vert\chi_l^m\right\vert\leq \sum_{j=0}^{2\Lambda}\sum_{|s|\leq j}\left\vert f_j^s\right\vert 
{{j}\choose{\left[\frac{j+1}{2}\right]}}\frac{1}{2^{\frac{j}{2}}}\left(e^{\frac{j\Lambda^2}{2k}}-1 \right)
\:\: R_j^s \:\: {\sum}'\sum_{0\leq h\leq
\min\left\{j,\frac{l+j-|m-s|}{2}\right\}}\left\vert\phi_{l+j-2h}^{m-s}\right\vert.
$$
Using
$$
\quad \left\vert f_j^s\right\vert\leq \|f\|,\quad \left\vert \phi_{l'}^{m'}\right\vert\leq \|\phi\|,\quad\min\left\{j,\frac{l+j-|m-s|}{2}\right\}\leq j\leq 2\Lambda
$$
and (\ref{differenceYlm}), (\ref{stime2})  we get
$$
\left\vert\chi_l^m \right\vert\leq  \left\| f\right\|\left\| \phi\right\|2\Lambda \sum_{j=1}^{2\Lambda}
T_j\, {{j}\choose{\left[\frac{j+1}{2}\right]}}\frac{1}{2^{\frac{j}{2}}}\left(e^{\frac{j\Lambda^2}{2k}}-1 \right), \qquad\mbox{where }\: T_j:=\sum_{|s|\leq j}
\sqrt{\frac{(j+s)!2^{j-s}}{(2j)!(j-s)!}} j^{j-s}.
$$
Let us now prove that for all $j\ge 1$
\be
T_j\le 4 j^j, \qquad {{j}\choose{\left[\frac{j+1}{2}\right]}}\le 2^j.               \label{disug'}
\ee
By straightforward computations we find $T_1=3$, what fulfills (\ref{disug'}). For $j\ge 2$
we use the following inequality
\be
\sqrt{\frac{(j+s)!2^{j-s}}{(2j)!(j-s)!}}j^{j-s}\leq j^{\frac{j-s}{2}},           \label{disug}
\ee
which can be proved $\forall$ $j\in\mathbb{N}$ and $\forall$ $|s|\leq j$ iteratively. In fact, it is trivial when $s=j$. Moreover,
if it is true for some $s\in\,]\!-\!j,j]$, it is true also if we replace $s\rightarrow s-1$:
$$
\sqrt{\frac{(j+s-1)!2^{j-s+1}}{(2j)!(j-s+1)!}}j^{j-s+1}=\sqrt{\frac{(j+s)!2^{j-s}}{(2j)!(j-s)!}}j^{j-s}j\sqrt{\frac{2}{(j+s)(j-s+1)}}\leq j^{\frac{j-(s-1)}{2}}
$$
because $(j+s)(j-s+1)-2j=j(j-1)-s(s-1)\geq 0$.  (\ref{disug}) implies
$$
T_j=\sum_{|s|\leq j}\sqrt{\frac{(j+s)!2^{j-s}}{(2j)!(j-s)!}}j^{j-s}\leq
\sum_{|s|\leq j}j^{\frac{j-s}{2}}=\sum_{s=0}^{2j}\sqrt{j}^{s}=\frac{j^{\frac{2j+1}{2}}-1}{\sqrt{j}-1}
=\frac{j^j-j^{-1/2}}{1- j^{-1/2}}\le 4 j^j;
$$
the last inequality holds for $j\ge 2$. The proof  of the second inequality in (\ref{disug'}) 
is straightforward. 
Applying (\ref{disug'})   we find for all $|m|\le l\le \Lambda$
\bea
\left\vert\chi_l^m \right\vert \leq   \left\| f\right\|\left\| \phi\right\|8 \Lambda U_\Lambda,
\qquad \mbox{where } U_\Lambda:=\sum_{j=1}^{2\Lambda}
j^j 2^{\frac{j}{2}} \left(e^{\frac{j\Lambda^2}{2k}}-1 \right)\nn
Q_{\Lambda}:=\sum_{l=0}^{\Lambda}\sum_{|m|\leq l}\left\vert\chi_l^m \right\vert^2
\le  \left\| f\right\|^2\left\| \phi\right\|^2 64 \Lambda^2(\Lambda+1)^2\, U_\Lambda^2
.\label{abschi}
\eea
Setting $\sigma:=2^{3/2}\Lambda$, $\tau:=e^{\frac{\Lambda^2}{2k}}$ we find
\bea
U_\Lambda\le \sum_{j=1}^{2\Lambda}
(2\Lambda)^j 2^{\frac{j}{2}} \left(e^{\frac{j\Lambda^2}{2k}}-1 \right)=\sum_{j=1}^{2\Lambda}
\left(\sigma\tau \right)^j- \sum_{j=1}^{2\Lambda}\sigma^j
=\frac{\left(\sigma\tau \right)^{2\Lambda+1}-1}{\sigma\tau -1}-
\frac{\sigma^{2\Lambda+1}-1}{\sigma-1}\nn
=\frac{\sigma^{2\Lambda+2}\tau\left(\tau^{2\Lambda}-1\right)-
\sigma^{2\Lambda+1}\left(\tau^{2\Lambda+1}-1\right)+\sigma(\tau\!-\!1)}{(\sigma\tau -1)(\sigma-1)}
\le 2\frac{\sigma^{2\Lambda+2}\tau\left(\tau^{2\Lambda}-1\right)}{(\sigma\tau -1)(\sigma-1)}
\nonumber
\eea
For $\Lambda\ge 3$ we easily show
$$
\frac 1{\sigma-1}<\frac 1{2\Lambda},\quad\quad\frac 1{\sigma\tau-1}<\frac 1{2\Lambda\tau}
\qquad\Rightarrow\qquad U_\Lambda < \frac 12 \sigma^{2\Lambda}\left(\tau^{2\Lambda}-1\right)
=2^{3\Lambda-1}\Lambda^{\Lambda}\left(e^{\frac{\Lambda^3}{k}}-1\right).
$$
Replacing into (\ref{abschi}) we obtain
\bea
Q_{\Lambda}
\le  \left\| f\right\|^2\left\| \phi\right\|^2 (\Lambda+1)^2\, 2^{6\Lambda+4}\Lambda^{2\Lambda+2}\left(e^{\frac{\Lambda^3}{k}}-1\right)^2
< \left\| f\right\|^2\left\| \phi\right\|^2  
(\Lambda+1)^2\, 2^{6\Lambda+6}\frac{\Lambda^{2\Lambda+8}}{k^2}
.\label{abschi'}
\eea
The last inequality holds for sufficiently small $\Lambda^3/k$, e.g. $\Lambda^3/k<1/2$, because $e^x-1<2x$ for $0<x<1/2$. Finally, by (\ref{normffh}), (\ref{abschi'})
the choice $k(\Lambda)= 2^{3\Lambda+3}\Lambda^{\Lambda+5}(\Lambda\!+\!1)$ implies 
\bea
\Vert(f -\hat f_\Lambda)\phi\Vert^2\le \left\| f\right\|^2\left\| \phi\right\|^2  
\, \frac{1}{\Lambda^2}+\sum_{l>\Lambda}\sum_{|m|\leq l}|(f\phi)_l^m|^2 \:
\stackrel{\Lambda\to\infty}{\longrightarrow}0,       \label{flimit3D}
\eea
i.e. $\widehat{f}_\Lambda\to f\cdot$ strongly for all $f\in B\big(S^2\big)$, as claimed. 
Replacing $f\mapsto fg$,
 we find also that $\widehat{(fg)}_\Lambda\to (fg)\cdot$ strongly for all $f,g\in B\big(S^2\big)$.
On the other hand, relation (\ref{flimit3D})  implies also 
\bea
 \Vert(f -\hat f_\Lambda)\phi\Vert^2=\left\| f\right\|^2\left\| \phi\right\|^2  
\, \frac{1}{\Lambda^2}+\left\| f\phi\right\|^2\le \left(
\frac{\left\| f\right\|^2}{\Lambda^2}
+\left\| f\right\|_\infty^2\right)\left\| \phi\right\|^2, \nn[6pt]
\Vert \hat f_\Lambda\phi\Vert 
\le \Vert (\hat f_\Lambda\!-\! f)\phi\Vert +\Vert f\phi\Vert 
\le \Vert (\hat f_\Lambda\!-\! f)\phi\Vert +\Vert f\Vert_\infty \Vert\phi\Vert\le 
\left(\left\| f\right\|+2 \Vert f\Vert_\infty\right) \Vert\phi\Vert, \nonumber
\eea
i.e. the operator norms $\Vert \hat f_\Lambda\Vert_{op}$ of the $ \hat f_\Lambda$ are bounded  uniformly in
$\Lambda$: \ $\Vert \hat f_\Lambda\Vert_{op}\le \left\| f\right\|+2\Vert f\Vert_\infty$.
Therefore, as claimed, (\ref{flimit3D}) implies again also
\bea
\Vert(fg -\hat f_\Lambda\hat g_\Lambda)\phi\Vert
&\le & \Vert (f -\hat f_\Lambda)g\phi\Vert+\Vert \hat f_\Lambda(g-\hat g_\Lambda)\phi\Vert \nn[6pt]
&\le & \Vert (f -\hat f_\Lambda)(g\phi)\Vert+\Vert \hat f_\Lambda\Vert_{op} \: \: \Vert(g -\hat g_\Lambda)\phi\Vert
\stackrel{\Lambda\to\infty}{\longrightarrow}0.      \label{flimit'3D}
\eea

\end{document}